\documentclass[aps,prd,twocolumn,superscriptaddress]{revtex4-2}
\usepackage{graphicx}  
\usepackage{dcolumn}   
\usepackage{bm}        
\usepackage{amssymb}   
\usepackage{amsmath}   
\usepackage{bbm}
\usepackage{draftcopy}
\usepackage{relsize} 
\usepackage{xfrac}    
\usepackage{slashed}
\usepackage{color}
\usepackage{footnote}
\usepackage{array}
\definecolor{dgreen}{cmyk}{1.,0.,1.,0.2}        
\definecolor{orange}{cmyk}{0.,0.353,1.,0.}    

\usepackage[bookmarks]{hyperref}

\newcommand{\di}{{\rm d}}

\newcommand{\be}{\begin{equation}}
\newcommand{\ee}{\end{equation}}                                                                               
\newcommand{\bea}{\begin{eqnarray}}
\newcommand{\eea}{\end{eqnarray}} 

\graphicspath{{./figures/}}

\begin{document}
\title{Dilepton emission assists the search for the QCD critical point}
\author{Gaoqing Cao}
\email{caogaoqing@sysu.edu.cn}
\address{School of Physics and Astronomy, Sun Yat-sen University, Zhuhai 519088, China}
 \author{Xiaofeng Luo}
 \email{xfluo@ccnu.edu.cn}
 \address{Key Laboratory of Quark $\&$ Lepton Physics (MOE) and Institute of Particle Physics,
Central China Normal University, Wuhan 430079, China}
\author{Wei-jie Fu}
\email{wjfu@dlut.edu.cn}
\address{School of Physics, Dalian University of Technology, Dalian, 116024, China}
\date{\today}

\begin{abstract}
In this work, we propose that dilepton emission rate (DER) could possibly carry the characteristic structures associated with the chiral criticality of the QCD system based on the extended Polyakov-quark-meson model. The model could successfully capture two main mechanisms for dilepton production, $\pi^+\pi^-$ and quark-antiquark annihilations on one hand, and self-consistently account for chiral transition and (de-)confinement on the other hand. All the moments of the DER peak exhibit extremal features similar to those of light-quark mass along the chemical freeze-out lines, thus the DER can reflect the change of chiral symmetry and criticality. However, the DER is relatively too low to produce a sufficient number of thermal dileptons in each heavy-ion collision, which prevents the implementation of event-by-event measurements of the moments. As an alternative, we propose to study the baseline-subtracted DER: For a given low center of mass of dileptons, the reduced baseline-subtracted DER exhibits a nonmonotonic behavior in the region where the light-quark mass changes most rapidly.
\end{abstract}


\begin{titlepage}
\maketitle
\end{titlepage}

\emph{Introduction.} According to most recent Beam Energy Scan Phase-II at RHIC in USA~\cite{STAR:2021fge,STAR:2022etb,STAR:2025zdq}, the QCD matter is in the hadronic phase at the collision energy $\sqrt{s_{\rm NN}}=3\,{\rm GeV}$~\cite{STAR:2021fge,STAR:2022etb} and the baryon number kurtosis shows no critical peak at $\sqrt{s_{\rm NN}}=7.7\,{\rm GeV}$ and above~\cite{STAR:2025zdq}. It is now a broad consensus that the search of QCD critical point is at a ``critical point": whether or not a peak~\cite{Stephanov:2011pb, Fu:2021oaw, Fu:2023lcm, Lu:2025cls} would show up in between these collision energies will give a  definitive conclusion on whether the QCD critical point can be detected in relativistic heavy ion collisions (HICs). In the coming years, several other low-energy facilities would join the search, such as the FAIR in Germany, NICA in Russia and HIAF in China, making the present period the best and most important window to explore the QCD criticality~\cite{Luo:2017faz, Luo:2022mtp, Dupuis:2020fhh, Fu:2022gou}. In particular, the illumination of leptons would be greatly improved in CBM experiment, it is essential to investigate potential leptonic signals for the QCD critical point, such as dilepton emission~\cite{McLerran:1984ay,Rapp:1999ej}.

In principle, any observables sensitive to the QCD medium can serve as a probe of critical point, though the criteria may vary from case to case. It is well known that the most prominent feature of critical point is the strong fluctuations in its vicinity~\cite{Stephanov:2008qz}, so higher moments of these observables are in principle better candidates. How can one measure the higher moments of these observables in HICs? Take the moments defined with respect to the chemical potential $\mu_{\rm X}$ ($X$ is a conserved charge) for example, the expectation value of the observable $\hat{O}$ is given in a grand canonical ensemble~\cite{McLerran:1987pz} by
\bea
\langle\hat{O}\rangle\equiv {{\rm Tr}\, \hat{O}\, e^{-\hat{H}/T+\hat{N}_{\rm X}\mu_{\rm X}/T}/\ {\rm Tr}\, e^{-\hat{H}/T+\hat{N}_{\rm X}\mu_{\rm X}/T}},
\eea
and the $n$th order moments follow as
\bea
\!\!\!\!\!\!\!&& {\partial\langle\hat{O}\rangle\over\partial (\mu_{\rm X}/T)}=\langle \delta\hat{O}\delta \hat{N}_{\rm X}\rangle,\  {\partial^2\langle\hat{O}\rangle\over\partial (\mu_{\rm X}/T)^2}=\langle \delta\hat{O}(\delta \hat{N}_{\rm X})^2\rangle, \nonumber
\eea
\bea
\!\!\!\!\!\!\!&&{\partial^3\langle\hat{O}\rangle\over\partial (\mu_{\rm X}/T)^3}=\langle \delta\hat{O}(\delta \hat{N}_{\rm X})^3\rangle-3\langle \delta\hat{O}\delta \hat{N}_{\rm X}\rangle\langle (\delta \hat{N}_{\rm X})^2\rangle, \dots
\eea
with $\delta\hat{O}\equiv \hat{O}-\langle\hat{O}\rangle$.
Thus, the higher moments of the observables are nothing but just the correlations between their fluctuations and higher order fluctuations of the total conserved charge, $(\delta \hat{N}_{\rm X})^n$. This result can be easily extended to mixed moments when multiple conserved charges are involved, such as $X=B, Q, S$ in HICs~\cite{Luo:2017faz,Luo:2022mtp}. In most cases, the moments defined with respect to $\mu_{\rm B}$ are the most sensitive probes for the search of QCD critical point in $T-\mu_{\rm B}$ plane~\cite{Luo:2017faz,Luo:2022mtp}. Regarding the dilepton emission rate ${\cal R}$, derivatives with respect to $\mu_{\rm B}/T$ characterize its equilibrium response to baryon-number fluctuations and can be formally related to mixed correlations involving the baryon number within the grand-canonical framework.

In this work, we propose that dilepton emission rate in the low-mass region, $0.4 \, {\rm GeV}\leq M\leq 1.2  \, {\rm GeV}$, could inherit the characteristic structures associated with the chiral criticality of the QCD system. It is well known that dilepton emission receives contributions from many different sources, but recent experimental efforts have successfully isolated the thermal dilepton component, revealing that the temperature extracted from the low-mass region closely aligns with the chemical freeze-out line~\cite{STAR:2024bpc}. Motivated by this observation, we employ the freeze-out approximation throughout our calculations, despite the fact that the realistic fireball exhibits a spatially and temporally varying temperature profile. To carry out calculations, we extend the Polyakov-quark-meson (PQM) model~\cite{Schaefer:2009ui,Cao:2025zvh} by incorporating the vector sector~\cite{Pisarski:1995xu}, then the characteristic features of the chiral transition and (de-)confinement can be mapped onto the properties of dilepton emission. Along two freeze-out lines~\cite{Luo:2017faz}, the derivatives of the DER peak exhibit extremal features analogous to those of the light-quark mass, indicating that the DER inherits characteristic chiral-critical structures within the present equilibrium model. Because the low dilepton yield makes event-by-event measurements of these higher-order quantities impractical, we additionally examine the baseline-subtracted DER as a simpler rate-level diagnostic.

\emph{A chiral effective model.} Previously, the renormalizable $2+1$ flavor Polyakov-quark-meson model~\cite{Cao:2025zvh} has been successfully adopted to study the $T-\mu_{\rm B}$ phase diagram of QCD matter. The extracted pseudocritical temperature and the location of the QCD critical point are consistent with results from lattice QCD~\cite{Bellwied:2015rza,HotQCD:2018pds} and functional QCD approaches~\cite{Fu:2019hdw,Gao:2020fbl,Gunkel:2021oya}. Thus, it is convincing that the PQM model can effectively capture the QCD criticality and provide valuable insights into potential critical signals in heavy-ion collisions.‌ In Euclidean space with the metric $g^{\mu\nu}=-\delta_{\mu\nu}$, the original Lagrangian density of the PQM model~\cite{Schaefer:2009ui} is given by
\begin{eqnarray}
{{\cal L}_{\rm PQM}^0}&=&\bar{\psi}\left[i\gamma^\mu D_\mu\!-\! i\gamma^4{\mu_{\rm B}\over3}\!-\!g_{\rm sq}T^b\left(\sigma_b\!+\!i\gamma^5\pi_b\right)\right]\psi\nonumber\\
&&\!\!\!\!\!\!\!\!\!\!\!\!\!\!\!\!\!\!\!\!+{1\over2}\!\!\left[{\partial}_\mu \phi^{\dagger b}{\partial}^\mu \phi_b\!-\!m^2_\phi\phi^{\dagger b}\phi_b\right]\!-\!h_1[{\rm Tr}(\phi^\dagger\phi)]^2\!-\!h_2{\rm Tr}(\phi^\dagger\phi)^2\nonumber\\
&&\!\!\!\!\!\!\!\!\!\!\!\!\!\!\!\!\!\!\!\!+\kappa\, [{\rm Det}(\phi^\dagger)\!+\!{\rm Det}(\phi)]\!+\!c_b\, {\rm Tr}\ T^b(\phi^\dagger\!+\!\phi)\!-\!V(L),
\end{eqnarray}
where $\psi(x)=(u(x),d(x),s(x))^T$ denotes the three-flavor quark field with the charge matrix $Q_{q}={\rm diag}(q_{u}, q_{d}, q_{s})$, $D_\mu={\partial}_\mu+i\ Q_{q}A_\mu-ig_s{\cal A}^4$ is the covariant derivative with $A_\mu$ the electromagnetic (EM) field and ${\cal A}^4$ the temporal gluon field, and $\phi\equiv T^b(\sigma_b+i\,\pi_b)$ is the scalar-pseudoscalar field matrix. To study the dilepton emission which is itself a direct QED effect, we choose $\sigma_b+i\,\pi_b$ to be the eigenstates of electric charge with eigenvalues $q_{b}$. Correspondingly, the interaction matrices in flavor space are $ T^0=\sqrt{1\over6}{\bm 1}, T^{1,2}={\lambda^1\mp i\lambda^2\over2\sqrt{2}}, T^{4,5}={\lambda^4\mp i\lambda^5\over2\sqrt{2}}, T^b={\lambda^b\over2}\ (b=3,6,7,8)$ with  $\lambda^b$  the Gell-Mann matrices~\cite{Cao:2025zvh}. In the physical vacuum, $u$ and $d$ quarks are almost degenerate and chiral symmetry is broken through the scalar channels, thus only the components $b=0, 8$ of $c_b$ are supposed to be nonzero. 

The Polyakov loop $L\equiv {1\over N_{\rm c}}{\rm Tr}_{\rm c}e^{i\,g_s\int_0^{1/T}dx_4{\cal A}^4}$ determines when the quark degrees of freedom are dynamically relevant in the QCD matter. For simplicity, we would set $L=L^*$ even in the case with finite $\mu_{B}$ and take the empirical expression for the pure gauge potential $V(L)$ as given by Munich group~\cite{Ratti:2005jh}, 
\bea
{V(L)\over T^4}&=&-{a\left({T/ T_0}\right)\over2}L^2-{0.75\over3}L^3+{7.5\over4}L^4
\eea
with $a(x)=6.75-{1.95\over x}+{2.625\over x^{2}}-{7.44\over x^{3}}$. The feedbacks of quarks to gluons are also important and can be taken into account by modifying the deconfinement temperature $T_0$ according to the corrections of the two-loop $\beta$-function of QCD~\cite{Herbst:2010rf}, that is,
\bea
T_0\!=\!T_\tau\, e^{-{1/\alpha_{s}\over b}}, \ \ \ 
b\!=\!{33\!-\!2N_{f}\over6\pi}\!-\!{16N_{f}\over\pi}\!\!\left(\!{\mu_B\over 2T_\tau}\!\right)^2
\eea
with $T_\tau=1.77\ {\rm GeV}, \alpha_{s}=0.304$ and $N_{f}=3$. 

To accurately evaluate dilepton emission, the vector mesons must be taken into account to capture the nonperturbative features of QCD in the hadronic phase~\cite{Rapp:1999ej}. Specifically, the neutral vector mesons $\rho^0, \omega$ and $\phi$ are included in the calculation of dilepton production from charged pion annihilation within the vector dominance model~\cite{Gounaris:1968mw,Gale:1990pn}. Among these channels, $\rho^0$ is identified as the most sensitive probe for studying the approximate $SU(2)$ chiral symmetry breaking and restoration, since it is not chirally invariant and predominantly decays into $\pi^+\pi^-$. In this work, we focus on the evolution features of dilepton emission via the $\rho^0$ channel in the hadronic phase. The lowest-lying vector states $\rho^0, \rho^\pm$ and axial-vector states $a_1^0,a_1^\pm$ are introduced to preserve the approximate $SU(2)$ chiral symmetry of the extended model~\cite{Pisarski:1995xu}. Then, the corresponding corrections to the Lagrangian from these vector mesons are~\cite{Gale:1990pn,Pisarski:1995xu}
\begin{eqnarray}
{\Delta{\cal L}_{\rm PQM}}\!&=&\!-{1\over4} \left(\rho^b_{\mu\nu}\rho_{b}^{\mu\nu}\!+\!a_{1\mu\nu}^ba_{1b}^{\mu\nu}\right)\!+\!{m_{\rho}^2\over2}\left(\rho^b_{\mu}\rho_{b}^{\mu}\!+\!a_{1\mu}^ba_{1b}^{\mu}\right)\nonumber\\
&&+\bar{\psi}\left[g_{\rm vq}T^b\gamma^\mu (\rho_{b \mu}+i\,\gamma^5 a_{1b \mu})\right]\psi+{\cal L}_{\rm sv},
\end{eqnarray}
where the two terms in the first line correspond to the kinetic part and the mass part, respectively, and the first term in the second line describes quark-(pseudo-)vector meson interactions. The scalar-vector interaction term ${\cal L}_{\rm sv}$ can be introduced by performing the substitutions in the lightest flavor sector~\cite{Pisarski:1995xu}, ${\partial}_{\mu}\sigma_{\rm l}\rightarrow ({\cal D}^\mu \sigma)_{\rm l}\equiv {\partial}_{\mu}\sigma_{\rm l}-g_{\rm sv}a_{1 \mu}^{b\dagger}\pi_b$ and $\partial^\mu \pi^b\rightarrow ({\cal D}^\mu \pi)^b\equiv \partial_\mu\pi^b+g_{\rm sv}(i\varepsilon^{bcd}\pi_c^\dagger\rho_{d\mu}^\dagger+a_{1 \mu}^b\sigma_{\rm l})$ with $b,c,d=1,2,3$ corresponding to the charges $+,-,0$. 

\emph{Dilepton emission rate.}  To relate the QCD-related signal to the dilepton emission rate, the QED theory for leptons is required, and the corresponding Lagrangian is well established,
\bea
{\cal L}_{\rm QED}\!=\!-{1\over4}F_{\mu\nu}F^{\mu\nu}\!+\!\sum_{l=e,\mu}\!\bar{l}\left[i\gamma^\mu({\partial}_\mu\!+\!i\, eA_\mu)-m_{l}\right]l
\eea
with the EM strength tensor $F_{\mu\nu}\equiv\partial_\mu A_\nu-\partial_\nu A_\mu$ and  the light lepton field $l=e,\mu$. Eventually, by coupling the established chiral effective model to the QED theory through the $\rho^0$-photon conversion vertex~\cite{Rapp:1999ej}, the total Lagrangian can be summarized as
\bea
{\cal L}={{\cal L}_{\rm PQM}^0}+{\Delta{\cal L}_{\rm PQM}}+{\cal L}_{\rm QED}-{e\over g_{\rm sv}} m_{\rho}^{2} \rho_{\mu}^0A^\mu.
\eea

\begin{figure}[!htb]
	\begin{center}
		\includegraphics[width=8cm]{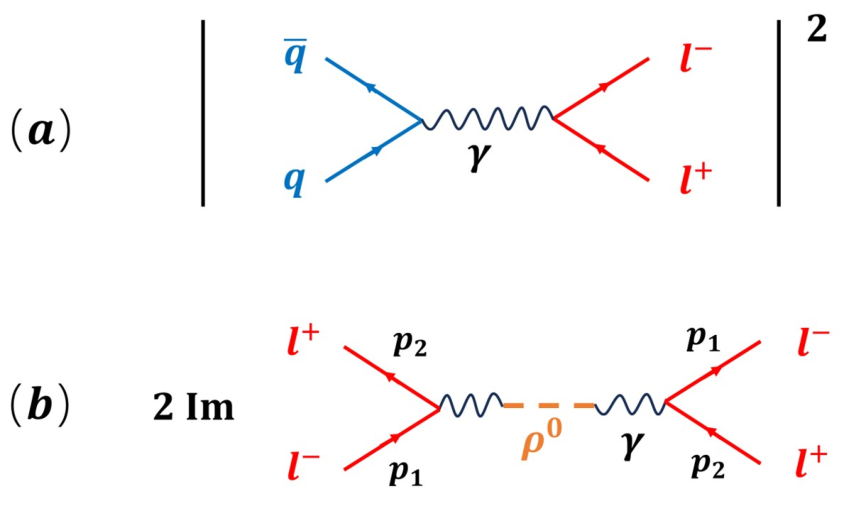}
		\caption{The main dilepton production mechanisms in QCD matter: (a) direct quark-antiquark annihilations via QED, (b) indirect charged hadron annihilations described in terms of the imaginary part of the retarded $\rho^0$ meson propagator.}\label{mechanisms}
	\end{center}
\end{figure}

 In such a model, the DERs receive contributions from two distinct channels~\cite{McLerran:1984ay,Rapp:1999ej}: perturbative quark-antiquark annihilations within the QED framework and nonperturbative charged hadron annihilations via the $\rho^0$ current, see the illustrations of the corresponding Feynman diagrams in Fig.~\ref{mechanisms}.
According to Ref.~\cite{McLerran:1984ay}, the DER can be simply presented in terms of EM current correlation functions, that is,
\bea
\!\!\!\!\!\!\!{\di R\over\di^4 q}\!=\!-{\alpha^2\over6\pi^3}{L_{\rm l}\over M^2}\!\Bigg(\!{m_{\rho}^4\over  g_{\rm sv}^2}{2\ {\rm Im} (D^{\rm R}_{\rho^0} )^{\mu}_{\ \mu}\over e^{q^0/T}-1}\!+\!N_c\!\!\sum_{f=u,d,s}\!\!\!\!W^{\mu}_{({f}) \mu}\!\Bigg)\label{dRdq}
\eea
by following the vector dominance model~\cite{Gounaris:1968mw,Gale:1990pn} and accounting for the quark current contributions~\cite{Kajantie:1986dh}. Here, the simple function $L_{\rm l}(M^2)\equiv\sum_{l=e,\mu}\sqrt{1-{4\tilde{m}^2_{l}}}\left(1+{2\tilde{m}^2_{l}}\right)\  (\tilde{m}\equiv{m/\sqrt{M^2}})$ comes from the summations over spins and integrations over momenta of the final state dileptons~\cite{McLerran:1984ay}, and $D^{\rm R}_{\rho^0}(q)$ is the retarded $\rho^0$ meson propagator. Compared to the simple mesonic model~\cite{Gale:1990pn}, the retarded $\rho^0$ meson propagator $D^{\rm R}_{\rho^0}(q)$ contains extra contributions from quark loops. These additional terms can essentially account for the chiral transition effect on $\rho$ meson~\cite{Bazavov:2019www} and partially capture the contributions from charged baryon annihilations~\cite{Rapp:1999ej}, see Appendix.~\ref{PF} for its expression and the detailed derivations. A comment here: In general, the soft mode is expected to play a prominent role in the immediate vicinity of the QCD critical point. But only the extremely low-mass region, defined as $M\leq 0.2\, {\rm GeV}$, is significantly modified by the soft mode~\cite{Nishimura:2023not}. Thus, the soft mode is neglected here since our primary focus lies on  the region $0.4\, {\rm GeV}\leq M \leq 1.2\, {\rm GeV}$. 

The EM current correlation functions from quarks can be given explicitly as~\cite{McLerran:1984ay}
\bea
W^{\mu}_{(f) \mu}\!\!&=&\!\! -2\left({q_f\over e}\right)^2\!\int{\di^3{\bm k_1}\over(2
\pi)^3}{n_f^+\left(E_{\bm k_1}^f\right)\over 2E_{\bm k_1}^f}\!\int{\di^3{\bm k_2}\over(2
\pi)^3}{{n}_f^-\left(E_{\bm k_2}^f\right)\over 2E_{\bm k_2}^f}\nonumber\\
&&(2\pi)^4\delta^{(4)}(k_1+k_2-q) (k_1\cdot k_2+2m_f^2),\label{Wq}
\eea
where $E_{\bm k}^f=\sqrt{{\bm k}^2+m_f^2}$ are the quark or antiquark energies, and the corresponding distribution functions can be derived from the chiral effective model at mean field approximation as~\cite{Cao:2025zvh}
\bea
n_f^t(k)&=&{L\, H_f^t+2L\,(H_f^t)^2+(H_f^t)^3\over 1+3L\, H_f^{t}+3L\,(H_f^{t})^2+(H_f^{t})^3}\label{nf}
\eea
with the auxiliary function $H_f^t(E_{\bm k}^f,\mu_{\rm B})\equiv e^{-{1\over T}\left(E_{\bm k}^f-t{\mu_{\rm B}\over 3}\right)}$. According to \eqref{nf}, the Polyakov loop $L$ governs the effective fermion degrees of freedom in QCD mater~\cite{Fukushima:2003fw}. In the confined phase with $L=0$, $n_f^t(p)$ reduces to the standard distribution functions for baryons and antibaryons; in the fully deconfined phase with $L=1$, $n_f^t(p)$ exactly recovers the distribution functions for quarks and antiquarks. Following the standard integral techniques provided in Ref.~\cite{Gale:1990pn}, we can get rid of the delta functions and angular variables in \eqref{Wq} to arrive at
\bea
\!\!\!\!\!\!\!\!\!\!\!W^{\mu}_{(f) \mu}
\!\!\!=\!\! -\!\left({q_f\over e}\right)^2\!\!\!\!\int_{k_-}^{k_+}\!\!\!{k\di{k}\over8
\pi}{n_f^+\!\!\left(E_{\bm k}^f\right)}{{n}_f^-\!\!\left(q_0\!-\!E_{\bm k}^f\right)}\!{M^2\!\!+\!\!2m_f^2\over  |{\bm q}|E_{\bm k}^f}\!\!
\eea
with the upper/lower integral limits defined as $k_{\pm}\equiv{ |{\bm q}|\over2}\left(1\pm\sqrt{1+(1-4\tilde{q}_0^2\tilde{m}_f^2)/{|\tilde{\bm q}|^2}}\right)$.

\emph{The dilepton responses to QCD critical point.} To carry out numerical calculations, we fix the parameters of PQM model as~\cite{Cao:2025zvh}
\bea
\!\!\!\!\!\!\!&&g_{\rm sq}=5.91, m_\phi=435\, {\rm MeV},
h_1=-2.70, h_2=46.5, \nonumber\\
\!\!\!\!\!\!\!&& \kappa=4810\, {\rm MeV}, c_0=(286\, {\rm MeV})^3, c_8=-(311\, {\rm MeV})^3, 
\eea
which gives the $\sigma_{\rm l}$ mass $m_{\sigma_{\rm l}}=500\,{\rm MeV}$, the pseudocritical temperature $T_{\rm pc}=159\, {\rm MeV}$ at $\mu_{\rm B}=0$, and the critical point $(\mu_{\rm Bc}, T_{\rm c})=(635,93)~{\rm MeV}$. For the vector sector, we take the vacuum mass of $\rho$ meson to be $m_\rho=770\, {\rm MeV}$, and fix $g_{\rm sv}=6.07$ and $g_{\rm vq}=1.55$ to reproduce the decay width $\Gamma_\rho=155\, {\rm MeV}$ in the vacuum~\cite{Gale:1990pn} and the location of DER peak ($\sim m_\rho$) in high energy collision~\cite{Han:2024nzr}, respectively. To study the critical phenomena, we consider two realistic chemical freeze-out lines, $T=0.158-0.14 \mu_{\rm B}^2-0.04n\,\mu_{\rm B}^4\ \ (n=2,3)$ with the units ${\rm GeV}$ for both $T$ and $\mu_{\rm B}$~\cite{Luo:2017faz}. The numerical calculations are mainly divided into two steps:\\
 1. Based on the mean field approximation, self-consistently solve the quark masses, Polyakov loop, and pion mass from the coupled gap equations and two-body correlation function at given $T$ and $\mu_{\rm B}$.\\ 2. Input these precomputed physical quantities to evaluate the dilepton emission rate \eqref{dRdq} numerically.\\ In most relevant studies, researchers are more focused on the integrated dilepton emission rate~\cite{Rapp:1999ej}, such as the one integrating over the full three-momentum space,
\bea
{\cal R}\equiv{\di R\over\di M^2}=\int  {\di^3{\bm q}\over 2q^0}{\di R\over\di^4q}\Big|_{q_0\rightarrow \sqrt{M^2+\bm{q}^2}}.
\eea

\begin{figure}[!htb]
	\begin{center}
		\includegraphics[width=8cm]{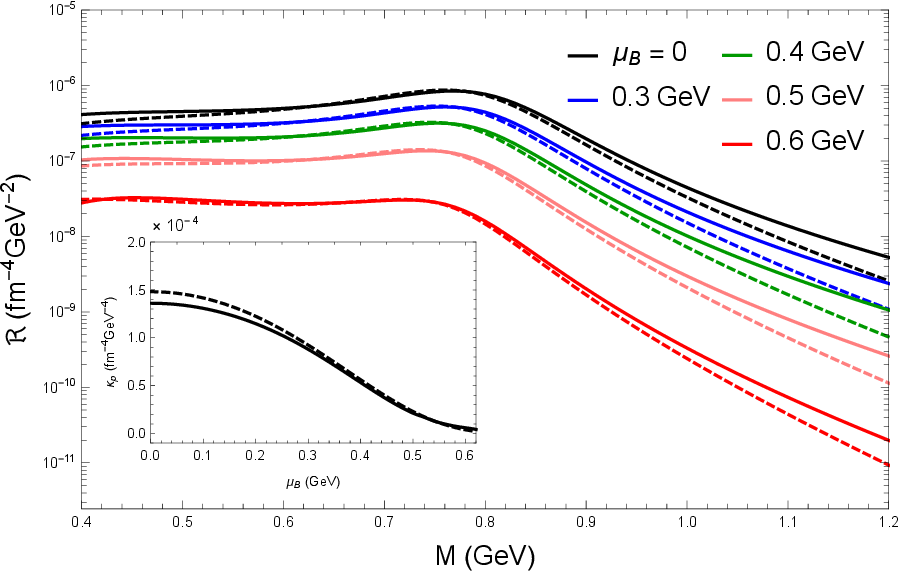}
		\caption{Along the $n=2$ freeze-out line, the integrated dilepton emission rates ${\cal R}$ as functions of the dilepton invariant mass $M$ for a series of fixed baryon chemical potential $\mu_{\rm B}$ values. The solid lines are full DERs receiving contributions from both Fig.~\ref{mechanisms} (a) and (b), and the dashed ones are the corresponding baselines ${\cal R}_0$ where only pions with fixed mass contribute~\cite{Gale:1990pn}. The insertion are the curvatures $\kappa_{\rm p}$ of the DER peaks as functions of the baryon chemical potential $\mu_{\rm B}$ for the full calculation (solid line) and baseline (dashed line).}\label{emrT}
	\end{center}
\end{figure}

Along the $n=2$ freeze-out line, which lies closer to the critical point, the integrated DERs ${\cal R}$ are illustrated in Fig.~\ref{emrT} for a series of fixed baryon chemical potential $\mu_{\rm B}$ values. Since the temperature decreases monotonically with $\mu_{\rm B}$ along the freeze-out line, the integrated DERs ${\cal R}$ also shows a decreasing trend as $\mu_{\rm B}$ rises.
Compared to the baseline distributions ${\cal R}_0$, the peaks of the full DER spectra are slightly shifted toward larger invariant mass $M$, which is consistent with the lattice result that the screening mass of $\rho$ meson increases after chiral symmetry restoration~\cite{Bazavov:2019www}. Obviously, the deviations between the full DERs and the corresponding baselines are more profound in the off-peak region than those in the peak region. The insertion shows that both the curvatures of the peak, $\kappa_{\rm p}\equiv \left|{\di^2{\cal R}\over\di M^2}\big|_{M_{\rm p}}\right|$, decrease with $\mu_{\rm B}$, consistent with the broadening phenomenon~\cite{Rapp:1999ej}. 

\begin{figure}[!htb]
	\begin{center}
		\includegraphics[width=8.5cm]{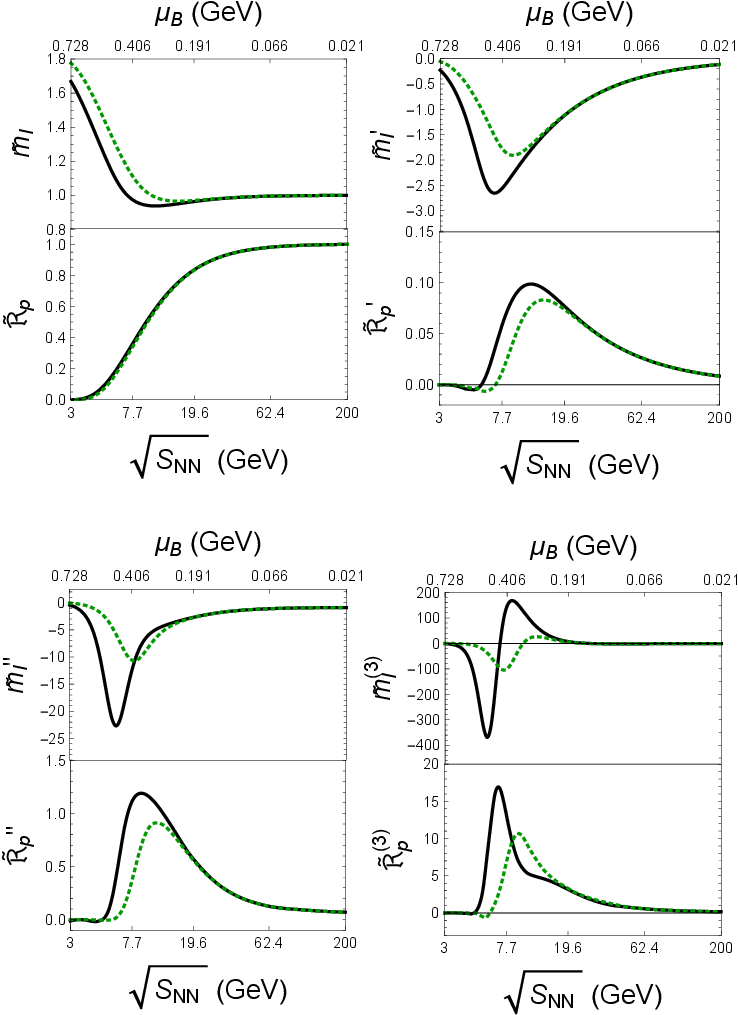}
		\caption{A comparison between the moments of the reduced light-quark mass, $\tilde{m}^{(n)}_{\rm l}$, and those of reduced dilepton emission rate, $\tilde{\cal R}^{(n)}_{\rm p}$, as functions of the baryon chemical potential $\mu_{\rm B}$ and colliding energy $\sqrt{S_{NN}}$. Here, the black solid and green dashed lines correspond to the freeze-out lines $n=2$ and $n=3$, respectively.}\label{mR}
	\end{center}
\end{figure}

In the following, we take the peak value of the DER, defined as ${\cal R}_{\rm p}\equiv {\cal R}|_{M_{\rm p}}$, as an illustrative example to show that the DER could effectively capture the features of chiral criticality. For this purpose, we introduce dimensionless moments of the reduced light-quark mass $\tilde{m}_{\rm l}\equiv m_{\rm l}/m_{\rm l}|_{\mu_{\rm B}=0}$ and the reduced DER $\tilde{\cal R}_{\rm p}\equiv{\cal R}_{\rm p}/{\cal R}_{\rm p}|_{\mu_{\rm B}=0}$ as
\bea
\tilde{m}^{(n)}_{\rm l}\equiv {\partial^n \tilde{m}_{\rm l}\over\partial (\mu_{\rm B}/T)^n}, \ \ \ \tilde{\cal R}_{\rm p}^{(n)}\equiv {\partial^n \tilde{\cal R}_{\rm p}\over\partial (\mu_{\rm B}/T)^n}
\eea
with the temperature fixed at the freeze-out point. The results are demonstrated altogether in Fig.~\ref{mR} for the moments with $n=0,1,2,3$ as functions of both $\mu_{\rm B}$ and the corresponding colliding energy $\sqrt{s_{NN}}$~\cite{Luo:2017faz}.

As one can see, up to a sign, the features of $\tilde{\cal R}_{\rm p}^{(n)}$ closely mirror those of $\tilde{m}^{(n)}_{\rm l}$; particularly for larger values of $n$, the locations of the extrema align very well with each other. Thus, within the present model, the DER reflects the evolution of chiral symmetry and the associated critical structure. Since the order parameter ${m}_{\rm l}$ is related to the first derivative of the pressure, $S_m\sigma_m\equiv \tilde{m}^{(3)}_{\rm l}/\tilde{m}'_{\rm l}$ and $S_R\sigma_R\equiv \tilde{\cal R}_{\rm p}^{(3)}/\tilde{\cal R}'_{\rm p}$ effectively correspond to $\kappa\sigma^2$ for baryon number fluctuations. According to our calculations, the combination $S_R\sigma_R$ is formally analogous to $\kappa\sigma^2$ in its derivative structure~\cite{Luo:2017faz,Fu:2019hdw}. However, since it is constructed from derivatives of an equilibrium emission rate rather than an experimentally measurable dilepton cumulant, its magnitude should not be interpreted as a direct measure of experimental sensitivity relative to baryon cumulants. Moreover, when the freeze-out line is shifted from $n=2$ to $n=3$, the kink in $\tilde{\cal R}_{\rm p}^{(3)}$ around $\mu_{\rm B}\sim 0.3\,{\rm GeV}$ becomes essentially suppressed, indicating that the DER is sensitive to the relative location of the critical point.

\begin{figure}[!htb]
	\begin{center}
		\includegraphics[width=8.5cm]{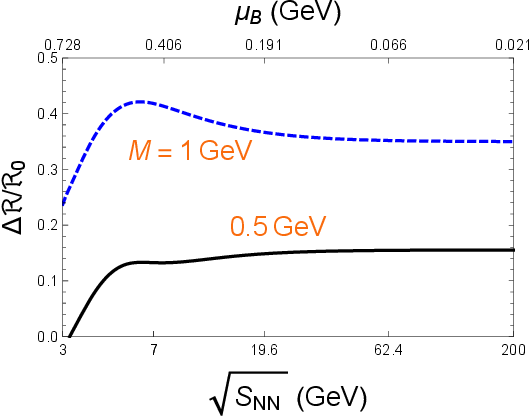}
		\caption{The relative deviations of the full DERs from the corresponding baselines, $\Delta {\cal R}/{\cal R}_0$, as functions of the baryon chemical potential $\mu_{\rm B}$ and colliding energy $\sqrt{S_{NN}}$ along the $n=2$ freeze-out line, for two selected invariant mass values $M=0.5\,{\rm GeV}$ (black solid) and $M=1\,{\rm GeV}$ (blue dashed).}\label{dRmu}
	\end{center}
\end{figure}

Nevertheless, the above discussions have not accounted for the finite small volume of each individual fireball produced in heavy-ion collisions. As a result, the thermal dilepton yield per event is small, making high-precision event-by-event measurements of the higher-order quantities considered above impractical. Instead, we can achieve better constraints on the thermal DER spectra at fixed freeze-out points, such as the cases presented in Fig.~\ref{emrT}. Using the fixed-mass pion result ${\cal R}_0$ as a reference, we examine $\Delta {\cal R}/{\cal R}_0$ to characterize how the full model prediction departs from this baseline along the freeze-out line, see Fig.~\ref{dRmu} for two values of  $M$. As clearly shown, within the present model, the nonmonotonic behavior of $\Delta {\cal R}/{\cal R}_0$ correlates with the region where the light-quark mass changes most rapidly and provides an additional rate-level manifestation of the modeled chiral-critical response.

\emph{Summary and discussions.} In this work, we propose that dilepton emission could inherit the characteristic structures associated with the chiral criticality of the QCD system. To demonstrate this, the well-known Polyakov-quark-meson (PQM) model is extended to incorporate the vector interaction sector while preserving the approximate $SU(2)$ chiral symmetry. Such an extended model is capable of capturing two primary mechanisms for dilepton production, $\pi^\pm$ annihilations in the hadronic phases and quark-antiquark annihilations in the quark-gluon plasma. Within this model, the predicted pseudocritical temperature and the location of the critical point are in good agreement with the results from lattice QCD and functional QCD approaches. Consequently, the extended PQM model provides a credible framework for capturing the essential features of dilepton emission in the vicinity of the QCD critical region.

Within the freeze-out approximation, it is observed that the dilepton emission rate decreases progressively more drastically, and its spectral peak becomes increasingly flattened as the baryon chemical potential increases. Taking the peak value as an example, the higher derivatives exhibit extremal features similar to those of the light-quark mass, showing that the DER inherits the chiral-critical response within the present model. Because the corresponding event-by-event measurements are impractical, we further examine the baseline-subtracted rate $\Delta {\cal R}$ as an additional rate-level diagnostic. Whether these structures survive the full space-time evolution and appear in measurable dilepton spectra remains to be established. Future high-statistics dilepton measurements provide motivation for such dynamical studies.

We acknowledge that the current work constitutes only a preliminary step in the exploration of the QCD critical point using thermal dilepton emission as a probe. In actual heavy-ion collisions, the experimental conditions are far from the idealized assumptions adopted in this work. In principle, the thermal dilepton emission in the low-mass region receives simultaneous contributions from both the quark-gluon plasma phase and the subsequent hadronization process, and the emission rate is spatially inhomogeneous across the entire fireball. To properly address these remaining issues and establish a direct connection with experiment, relativistic hydrodynamics can be employed to systematically investigate how the full space-time evolution of the hot medium modifies the dilepton emission spectrum.

\emph{Acknowledgement} G.C. thanks C. Yang for helpful discussions on many occasions. The work is funded by the Forefront Leading Project of Theoretical Physics from the National Natural Science Foundation of China with Grant No. 12447102. G.C. is also funded by the National Natural Science Foundation of China with Grant No. 12575152 and the Natural Science Foundation of Guangdong Province with Grant No. 2024A1515011225. X.F.L. is also supported by the National Natural Science Foundation of China with Grant Nos. 12122505, 11890711, and the National Key Research and Development Program of China with contract Nos. 2022YFA1604900, 2020YFE0202002, 2018YFE0205201. W.J.F. is also supported by the National Natural Science Foundation of China with Grant No. 12175030.

\appendix
\begin{widetext}
\section{The dressed propagator of $\rho^0$ meson}\label{PF}

In the extended Polyakov-quark-meson model, the propagator of $\rho^0$ meson receives corrections from both charged pion and quark loops as demonstrated in Fig.~\ref{rhop}.
\begin{figure}[!htb]
	\begin{center}
		\includegraphics[width=8cm]{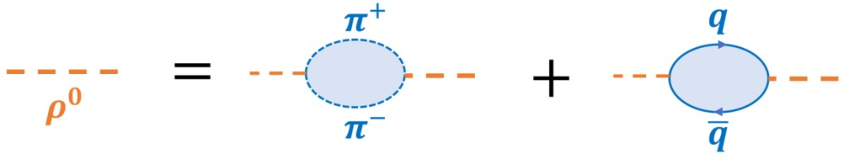}
		\caption{The self-consistent Feynman diagrams for the dressed propagator of $\rho^0$ meson (thick dashed line) in terms of its bare propagator (thin dashed line), charged pion loop (dotted loop) and quark loops (arrowed solid loop). }\label{rhop}
	\end{center}
\end{figure}
At one-loop approximation, the Feynman diagrams can be translated into mathematical expression for the inverse dressed propagator of $\rho^0$ meson, $\left({\cal D}_{ \rho^{0}}^{-1}\right)^{\mu\nu}({q})$, as
\bea
\left({\cal D}_{ \rho^{0}}^{-1}\right)^{\mu\nu}({q})=(q^2-{m}_{\rho}^2)g^{\mu\nu}-q^\mu q^\nu+\Pi_{\rm \rho^0}^{\mu\nu}({q})=\left[F(q)-q^2\right]{\cal P}_{\rm L}^{\mu\nu}+\left[G(q)-q^2\right]{\cal P}_t^{\mu\nu}-{m}_{\rho}^2g^{\mu\nu}.\label{rho0}
\eea
Here, the self-energy $\Pi_{\rm \rho^0}^{\mu\nu}({q})$ are composed of the contributions from charged pion loops $\Pi_{\rm \rho^0\pi}^{\mu\nu}({q})$ and quark loops $\Pi_{\rm \rho^0q}^{\mu\nu}({q})$, that is, $\Pi_{\rm \rho^0}^{\mu\nu}({q})=\Pi_{\rm \rho^0\pi}^{\mu\nu}({q})+\Pi_{\rm \rho^0q}^{\mu\nu}({q})$. Due to the violation of Lorentz invariance but intact of spatial rotational invariance at finite temperature and baryon chemical potential, the propagator is further separated into transverse and longitudinal polarized parts in the second step with the polarization tensors~\cite{Gale:1990pn}
\bea
{\cal P}_t^{00}={\cal P}_t^{\rm 0i}={\cal P}_t^{\rm i0}=0,\ {\cal P}_t^{ij}=-(g^{ij}+q^iq^j/{\bm q}^2),\ {\cal P}_{\rm L}^{\mu\nu}=q^{\mu}q^{\nu}/{q}^2-g^{\mu\nu}-{\cal P}_t^{\mu\nu}\ \ (i,j=1,2,3).
\eea
The auxiliary functions are given by $F(q)=({q}^2/{\bm q}^2)\Pi_{\rm \rho^0}^{00}$ and $G(q)={\cal P}_{\rm T\, ij}\Pi_{\rm \rho^0}^{ij}/2$, which, similar to the self-energy, also receive contributions from both charged pion loops and quark loops, so we can separate them as $F(q)=F_\pi(q)+F_{\rm q}(q)$ and $G(q)=G_\pi(q)+G_{\rm q}(q)$. Then, the dressed $\rho^0$ propagator can be derived as~\cite{Gale:1990pn}
\bea
{\cal D}_{ \rho^{0}}^{\mu\nu}({q})=-{{\cal P}_{\rm L}^{\mu\nu}\over q^2-m_\rho^2-F(q)}-{{\cal P}_t^{\mu\nu}\over q^2-m_\rho^2-G(q)}-{q^{\mu}q^{\nu}\over m_\rho^2q^2},
\eea
and the retarded one is given by $\left({\cal D}^{\rm R}_{\rho^0} \right)^{\mu\nu}(q)\equiv {\cal D}_{ \rho^{0}}^{\mu\nu}({q})|_{q_0\rightarrow E+i\,0^+}$ with $E$ a positive energy. Eventually, the imaginary part of the retarded propagator, which is directly related to dilepton production, can be evaluated after taking trace over the Lorentz indices as
\bea
{\rm Im} \left({\cal D}^{\rm R}_{\rho^0} \right)^{\mu}_{\ \mu}(q)={{\rm Im}\, F(q)\over \left[q^2-m_\rho^2-{\rm Re}\, F(q)\right]^2+\left[{\rm Im} \,F(q)\right]^2}+{2\,{\rm Im} \,G(q)\over \left[q^2-m_\rho^2-{\rm Re}\, G(q)\right]^2+\left[{\rm Im}\, G(q)\right]^2}.
\eea

The auxiliary functions from pion loops had been calculated in Ref.~\cite{Gale:1990pn}, where they were decomposed into vacuum and thermal parts, $F_\pi(q)= F_\pi^{\rm v}(q)+F_\pi^t(q)$ and $G_\pi(q)= G_\pi^{\rm v}(q)+G_\pi^t(q)$. The involved functions are
\bea
\!\!\!\!\!\!\!\!F_\pi^{\rm v}\!\!&=&\!\!G_\pi^{\rm v}\!=\!{g_{\rm sv}^2M^2\over 48\pi^2}\!\!\!\left[(1\!-\!4\tilde{m}^2_{\pi})^{3\over 2}\!\left(\!\ln\left|{\sqrt{1\!-\!4\tilde{m}^2_{\pi}}\!+\!1\over \sqrt{1\!-\!4\tilde{m}^2_{\pi}}\!-\!1}\right|\!-\!i\pi\, \theta(1\!-\!4\tilde{m}^2_{\pi})\!\right)\!+\!{8\tilde{m}^2_{\pi}}\!-\!{8{m}^2_{\pi}\over m_\rho^2}\!-\!\left(\!1\!-\!{4{m}^2_{\pi}\over m_\rho^2}\!\right)^{3\over 2}\!\ln\left|{\sqrt{1\!-\!{4{m}^2_{\pi}\over m_\rho^2}}\!+\!1\over \sqrt{1\!-\!{4{m}^2_{\pi}\over m_\rho^2}}\!-\!1}\right|\right],\\
\!\!\!\!\!\!\!\!F^t_\pi\!\!&=&\!\!{g_{\rm sv}^2M^2\over 4\pi^2|{\bm q}|^2}\int_0^\infty {k^2\di k\over E_\pi}{1\over e^{E_\pi/T}-1}\left[{4E_\pi^2+E^2\over 2k|{\bm q}|}(\ln|\alpha_\pi|-i\pi \Delta_\pi)+{2E_\pi E\over k|{\bm q}|}(\ln|\beta_\pi|+i\pi \Delta_\pi)-4\right],\label{Fm}\\
\!\!\!\!\!\!\!\!G^t_\pi\!\!&=&\!\!{g_{\rm sv}^2\over 4\pi^2}\int_0^\infty\! {k^2\di k\over E_\pi}{1\over e^{E_\pi/T}-1}\left[{4(|{\bm q}|^2k^2\!-\!E_\pi^2E^2)\!-\!M^4\over 4k|{\bm q}|^3}(\ln|\alpha_\pi|\!-\! i\pi \Delta_\pi)\!-\!{E_\pi EM^2\over k|{\bm q}|^3}(\ln|\beta_\pi|\!+\!i\pi \Delta_\pi)\!+\!{2M^2\over |{\bm q}|^2}\!+\!4\right]\label{Gm}
\eea
with $E_\pi=\sqrt{k^2+m_\pi^2}$, $E=\sqrt{ |{\bm q}|^2+M^2}$ and the auxiliary functions
\bea
&&\alpha_\pi\equiv{(M^2+2k|{\bm q}|)^2-4E_\pi^2E^2\over (M^2-2k|{\bm q}|)^2-4E_\pi^2E^2},\
\beta_\pi\equiv{M^4-4(k|{\bm q}|+E_\pi E)^2\over M^4-4(k|{\bm q}|-E_\pi E)^2},\\
&&\Delta_\pi={\rm Boole}(|E\sqrt{1\!-\!4\tilde{m}^2_{\pi}}- |{\bm q}||\leq 2k\leq E\sqrt{1\!-\!4\tilde{m}^2_{\pi}}+|{\bm q}|).
\eea
In the following, we will devote ourselves to deriving the self-energy $\Pi_{\rm \rho^0q}^{\mu\nu}({q})$ and the corresponding auxiliary functions, $F_{\rm q}(q)$ and $G_{\rm q}(q)$, from quark loops. 

\subsection{The self-energy from quark loops}
The self-energy from quark loops can be evaluated according to the second Feynman diagram on the right-hand side of Fig.~\ref{rhop} as
\bea
\Pi_{\rm \rho^0 q}^{\mu\nu}(q)=-{g_{v q}^2\over 4V_4}\sum_{f=u,d}^{c=r,g,b} {\rm tr}\ [G_{f,c}(k+q)\gamma^\mu G_{f,c}(k)\gamma^\nu],\label{PFrho}
\eea
where the traces ${\rm tr}$ are over Dirac matrices and energy momentum space, and the quark propagators are given at mean field approximation as~\cite{Cao:2025zvh}
\bea
G_{f,c}(k)={i\over -\slashed{k}- i\gamma^4\left(iq_{c}T+{\mu_{\rm B}\over3}\right)-m_f}\label{Gk}
\eea
with flavor indices ${f=u, d, s}$, color indices ${c=1, 2, 3}$, and $q_{c}$ the color charge. To account for effects of finite temperature and baryon chemical potential, it is simpler to work in Euclidean space. Then, after completing traces over Dirac matrices by utilizing the theorem ${\rm tr}\,\gamma^\mu\gamma^\rho\gamma^\nu\gamma^\sigma=4\left(g^{\mu\rho}g^{\nu\sigma}-g^{\mu\nu}g^{\rho\sigma}+g^{\mu\sigma}g^{\nu\rho}\right)$ and Matsubara frequency summation, the self-energy can be separated into a divergent vacuum term $\Pi_{ \rm \rho^0q(v)}^{\mu\nu}(q)$ and a convergent thermal term $\Pi_{ \rm \rho^0q(t)}^{\mu\nu}(q)$ with the forms
\bea
\Pi_{\rm \rho^0q(v)}^{\mu\nu}(q)\!&=&\! -{3g_{v q}^2\over 4}\sum_{f=u,d}\int{\di^4{ k}\over(2\pi)^4} {\rm tr}\!\!\left[{i\over -\slashed{k}-\slashed{q}-m_f}\gamma^\mu{i\over -\slashed{k}-m_f}\gamma^\nu\right],\\
\Pi_{ \rm \rho^0q(t)}^{\mu\nu}(q)\!&=&\!{g_{v q}^2\over 2}\sum_{f=u,d}^{c=r,g,b}\int{d^3{\bm k}\over(2\pi)^3}\sum_{u,t=\pm}{ 1\over E_{{\bm k}+u {{\bm q}\over 2}}^f\left[(E_{{\bm k}+u {{\bm q}\over 2}}^f+u\ t\ i q_4)^2-\left({\bm k}-u\ {{\bm q}\over 2}\right)^2-m_f^2\right]}{1\over 1+e^{\left[E_{{\bm k}+u {{\bm q}\over 2}}^f-t(i q_{c}T+ {\mu_{\rm B}\over3})\right]/T}}\nonumber\\
&&\!\!\!\!\!\!\!\!\!\left\{\left[-E_{{\bm k}+u {\bm q\over2}}^f(E_{{\bm k}+u {{\bm q}\over 2}}^f+u\ t\ i q_4)+{\bm k}^2-{{\bm q}^2\over 4}+m_f^2\right]g^{\mu\nu}+ \left[i\,t\,E_{{\bm k}+u {{\bm q}\over 2}}^fg^{4\mu}+\left({\bm k}+u {{\bm q}\over 2}\right)^{\mu}\right]\left[(i\,t\,E_{{\bm k}+u {{\bm q}\over 2}}^f-u\,q_4)g^{4\nu}\right.\right.\nonumber\\
&&\left.\left.+\left({\bm k}-u {{\bm q}\over 2}\right)^{\nu}\right]+ \left[i\,t\,E_{{\bm k}+u {{\bm q}\over 2}}^fg^{4\nu}+({\bm k}+u {{\bm q}\over 2})^{\nu}\right]\left[(i\,t\,E_{{\bm k}+u {{\bm q}\over 2}}^f-u\,q_4)g^{4\mu}+\left({\bm k}-u {{\bm q}\over 2}\right)^{\mu}\right]\right\}.
\eea
Here, the quark energies are defined as $E_{{\bm k}}^f\equiv\sqrt{{\bm k}^2+m_f^2}$.

One could immediately recognize that the vacuum term is exactly the same as that for photon polarization except the coupling constant. 
So by following the standard dimensional regularization in textbooks and applying the renormalization condition ${\rm Im}\, \Pi_{\rm \rho^0q(v)}^{\mu\nu}(q)=0$ and ${\rm Re}\,{\Pi_{\rm \rho^0q(v)}}{^{\mu}}_{\mu}|_{m_f\rightarrow m_{\rm f0}}^{q^2\rightarrow m_{\rho}^2}=0$, which means $\rho^0$ cannot decay into quark-antiquark pair for any $q^2$ and the self-energy vanishes at $q^2= m_{\rho}^2$ in vacuum, we have
\bea
\Pi_{\rm \rho^0q(v)}^{\mu\nu}(q)&=&-(q^2g^{\mu\nu}-q^\mu q^\nu){3g_{v q}^2\over 8\pi^2}\sum_{f=u,d}\int_0^1\di x\ x(1-x){\rm Re}\ln\left({m_f^2-x(1-x)q^2\over m_{\rm f0}^2-x(1-x)m_{\rho}^2}\right)\nonumber\\
&=&-(q^2g^{\mu\nu}-q^\mu q^\nu){g_{v q}^2\over 16\pi^2}\sum_{f=u,d}{\rm Re}\,\left[\ln {m_f^2m_{\rho}^2\over m_{\rm f0}^2q^2}-4\left({m_f^2\over q^2}-{m_{\rm f0}^2\over m_{\rho}^2}\right)+\sqrt{1-{4m_f^2\over q^2}}\left(1+{2m_f^2\over q^2}\right)\ln{\sqrt{1-{4m_f^2\over q^2}}+1\over \sqrt{1-{4m_f^2\over q^2}}-1}\right.\nonumber\\
&&\qquad\qquad\qquad\left.-\sqrt{1-{4m_{\rm f0}^2\over m_{\rho}^2}}\left(1+{2m_{\rm f0}^2\over m_{\rho}^2}\right)\ln{\sqrt{1-{4m_{\rm f0}^2\over m_{\rho}^2}}+1\over \sqrt{1-{4m_{\rm f0}^2\over m_{\rho}^2}}-1}\right].
\eea
In Minkowski space, the contributions to the self-energies of longitudinal and transverse polarized $\rho^0$ mesons are both
\bea
F_{\rm q}^{\rm v}&=&G_{\rm q}^{\rm v}={g_{v q}^2M^2\over 16\pi^2}\sum_{f=u,d}{\rm Re}\,\left[\ln {\tilde{m}_f^2m_{\rho}^2\over {m}_{\rm f0}^2}-4(\tilde{m}_f^2-{{m}_{\rm f0}^2\over m_{\rho}^2})+\sqrt{1-{4\tilde{m}_f^2}}\left(1+{2\tilde{m}_f^2}\right)\ln{\sqrt{1-{4\tilde{m}_f^2}}+1\over \sqrt{1-{4\tilde{m}_f^2}}-1}\right.\nonumber\\
&&\qquad\qquad\qquad\left.-\sqrt{1-{4m_{\rm f0}^2\over m_{\rho}^2}}\left(1+{2m_{\rm f0}^2\over m_{\rho}^2}\right)\ln{\sqrt{1-{4m_{\rm f0}^2\over m_{\rho}^2}}+1\over \sqrt{1-{4m_{\rm f0}^2\over m_{\rho}^2}}-1}\right].
\eea

\subsection{The thermal part of the self-energy from quark loops}\label{qloop}
Now, we focus on the thermal term which can be reduced to a more compact form as the following
\bea
\Pi_{ \rm \rho^0q(t)}^{\mu\nu}(q)&=&{3g_{v q}^2\over2}\sum_{f=u,d}\int{d^3{\bm k}\over(2\pi)^3}\sum_{u,t=\pm}{ n_f^t\left({\bm k}+{{\bm q}\over 2}\right)\over E_{{\bm k}+{{\bm q}\over 2}}^f[2u\ t\ E_{{\bm k}+{{\bm q}\over 2}}^f( i q_4)+2{\bm q}\cdot{\bm k}+(i q_4)^2]}\left\{\left[-u\ t\ E_{{\bm k}+{{\bm q}\over 2}}^f( i q_4)-{\bm q}\cdot{\bm k}-{{\bm q}^2\over2}\right]g^{\mu\nu}\right.\nonumber\\
&&\left.+2 \left[(i\,u\,t\,E_{{\bm k}+{{\bm q}\over 2}}^f-{q_4\over2})g^{4\mu}+{\bm k}^\mu\right]\left[(i\,u\,t\,E_{{\bm k}+{{\bm q}\over 2}}^f-{q_4\over2})g^{4\nu}+{\bm k}^\nu\right]-{q^\mu q^\nu\over2}\right\}\nonumber\\
&=&-{3g_{v q}^2\over2}\left[g^{\mu\nu}+\left(1-{{\bm q}^2\over q_4^2}\right)g^{4\mu}g^{4\nu}+g^{4\mu}{{\bm q}^\nu\over q_4}+g^{4\nu}{{\bm q}^\mu\over q_4}\right]\sum_{f=u,d}\int{d^3{\bm k}\over(2\pi)^3}\sum_{t=\pm}{n_f^{t}({\bm k}+{{\bm q}\over 2})\over E_{{\bm k}+{{\bm q}\over 2}}^f}\nonumber\\
&&+{3g_{v q}^2\over4}\sum_{f=u,d}\int{d^3{\bm k}\over(2\pi)^3}\sum_{u,t=\pm}{ q^2g^{\mu\nu}-q^\mu q^\nu+4\left({{\bm q\cdot k}\over q_4}g^{4\mu}- {\bm k}^\mu\right)\left({{\bm q\cdot k}\over q_4}g^{4\nu}- {\bm k}^\nu\right)\over E_{{\bm k}+{{\bm q}\over 2}}^f[2u\ t\ E_{{\bm k}+{{\bm q}\over 2}}^f( i q_4)+2{\bm q}\cdot{\bm k}+(i q_4)^2]}n_f^t\left({\bm k}+{{\bm q}\over 2}\right)\nonumber\\
&=&-{3g_{v q}^2\over2}\left[g^{\mu\nu}+\left(1-{{\bm q}^2\over q_4^2}\right)g^{4\mu}g^{4\nu}+g^{4\mu}{{\bm q}^\nu\over q_4}+g^{4\nu}{{\bm q}^\mu\over q_4}\right]\sum_{f=u,d}\int{d^3{\bm k}\over(2\pi)^3}\sum_{t=\pm}{n_f^{t}({\bm k})\over E_{{\bm k}}^f}\nonumber\\
&&+{3g_{v q}^2\over4}\sum_{f=u,d}\int{d^3{\bm k}\over(2\pi)^3}\sum_{u,t=\pm}{ q^2g^{\mu\nu}\!-\!q^\mu q^\nu\!+\!\left({2{\bm q\cdot k}-{\bm q}^2\over q_4}g^{4\mu}\!-\!2{\bm k}^\mu+{\bm q}^\mu\right)\left({2{\bm q\cdot k}-{\bm q}^2\over q_4}g^{4\nu}\!-\!2{\bm k}^\nu+{\bm q}^\nu\right)\over [2u\ t\ E_{{\bm k}}^f( i q_4)+2{\bm q}\cdot{\bm k}+q^2]}{n_f^{t}({\bm k})\over E_{{\bm k}}^f}.\label{Pit}
\eea
During the derivations, the tricks utilized are as follows: In the first step, the color degree of freedom is summed over to give rise to the particle distribution functions $n_f^t\left({\bm k}\right)\equiv \left[LH_f^{t}+2L(H_f^{t})^2+(H_f^{t})^3\right]\left[1+3L\, H_f^{t}+3L\,(H_f^{t})^2+(H_f^{t})^3\right]^{-1}$ with $H_f^t(E_{\bm k}^f,\mu_{\rm B})\equiv e^{-{1\over T}\left(E_{\bm k}^f-t{\mu_{\rm B}\over 3}\right)}$, and the momentum has been changed from ${\bm k}$ to $u\ {\bm k}$ in the integrand. In the second step, we try to simplify the formula by getting rid of the energy term $E_{{\bm k}+{{\bm q}\over 2}}^f$ in the numerator. In the last step, we shift ${\bm k}+{{\bm q}\over 2}$ to ${\bm k}$ in the integrands in order to use odd-evenness to reduce the formula later. 

In the following, we will try to separate the longitudinal part from the transverse part and then complete the integrations over solid angles to facilitate numerical calculations. Firstly, the most central component of the longitudinal part is the temporal diagonal term, $\Pi_{ \rm \rho^0q(t)}^{\rm 44}(q)$, which can be further reduced from \eqref{Pit} as
\bea
\Pi_{ \rm \rho^0q(t)}^{\rm 44}(q)&=&{3g_{v q}^2\over2}{{\bm q}^2\over q_4^2}\sum_{f=u,d}\int{d^3{\bm k}\over(2\pi)^3}\sum_{t=\pm}{n_f^{t}({\bm k})\over E_{{\bm k}}^f}+{3g_{v q}^2\over4}\sum_{f=u,d}\int{d^3{\bm k}\over(2\pi)^3}\sum_{u,t=\pm}{ {\bm q}^2+\left({2{\bm q\cdot k}-{\bm q}^2\over q_4}\right)^2\over [2u\ t\ E_{{\bm k}}^f( i q_4)+2{\bm q}\cdot{\bm k}+q^2]}{n_f^{t}({\bm k})\over E_{{\bm k}}^f}\nonumber\\
&=&{3g_{v q}^2\over2}\sum_{f=u,d}\int{d^3{\bm k}\over(2\pi)^3}\sum_{t=\pm}{n_f^{t}({\bm k})\over E_{{\bm k}}^f}+{3g_{v q}^2\over4}\sum_{f=u,d}\int{d^3{\bm k}\over(2\pi)^3}\sum_{u,t=\pm}{ {\bm q}^2+\left(q_4-2i\,u\ t\ E_{{\bm k}}^f\right)^2\over [2u\ t\ E_{{\bm k}}^f( i q_4)+2{\bm q}\cdot{\bm k}+q^2]}{n_f^{t}({\bm k})\over E_{{\bm k}}^f}\nonumber\\
&=&{3g_{v q}^2\over2}\sum_{f=u,d}\int{d^3{\bm k}\over(2\pi)^3}\sum_{t=\pm}{n_f^{t}({\bm k})\over E_{{\bm k}}^f}-{3g_{v q}^2\over4}\sum_{f=u,d}\int{d^3{\bm k}\over(2\pi)^3}\sum_{u,t=\pm}{ 4\left(E_{{\bm k}}^f\right)^2+q^2+4u\ t\ E_{{\bm k}}^f( i q_4)\over [2u\ t\ E_{{\bm k}}^f( i q_4)+2{\bm q}\cdot{\bm k}+q^2]}{n_f^{t}({\bm k})\over E_{{\bm k}}^f}\nonumber\\
&=&-{3g_{v q}^2\over8}\sum_{f=u,d}\int{k\di k\over(2\pi)^2|{\bm q}|}\sum_{t=\pm}\left[-8|{\bm q}|k+\left(4\left(E_{{\bm k}}^f\right)^2+q^2\right)\ln \alpha_f^{\rm E}+4 E_{{\bm k}}^f( i q_4)\ln \beta_f^{\rm E}\right]{n_f^{t}({\bm k})\over E_{{\bm k}}^f},\\
&&\alpha_f^{\rm E}\equiv {(2|{\bm q}|k+q^2)^2+4\left(E_{{\bm k}}^f\right)^2q_4^2\over (2|{\bm q}|k-q^2)^2+4\left(E_{{\bm k}}^f\right)^2q_4^2},\ \ \ \ \  \beta_f^{\rm E}\equiv {4(|{\bm q}|k+E_{{\bm k}}^f( i q_4))^2-q^4\over 4(|{\bm q}|k-E_{{\bm k}}^f( i q_4))^2-q^4}.\label{Pi44}
\eea
In the second step, we have gotten rid of the term ${\bm q}\cdot{\bm k}$ in the numerator to facilitate the integration over the polar angle between ${\bm q}$ and ${\bm k}$; in the last step, we have completed both the integrations over solid angles and summation over $u$. 

Secondly, according to Ref.~\cite{Gale:1990pn}, the other nonvanishing components of the longitudinal part can be given by $\Pi_{\rm \rho^0q(t)}^{4i}(q)=-{q^4q^i\over {\bm q}^2}\Pi_{\rm \rho^0q(t)}^{\rm 44}(q)\ (i=1,2,3)$. The relation can be easily justified for the first term of \eqref{Pit}, now we are going to show that it is also true for the second term. If we define the tensor in the numerator of the integrand in the second term of \eqref{Pit} by
\bea
&&B^{\mu\nu}\equiv q^2g^{\mu\nu}\!-\!q^\mu q^\nu\!+\!\left({2{\bm q\cdot k}-{\bm q}^2\over q_4}g^{4\mu}\!-\!2{\bm k}^\mu+{\bm q}^\mu\right)\left({2{\bm q\cdot k}-{\bm q}^2\over q_4}g^{4\nu}\!-\!2{\bm k}^\nu+{\bm q}^\nu\right),
\eea
it follows that
\bea
&&B^{44}={\bm q}^2+\left({2{\bm q\cdot k}-{\bm q}^2\over q_4}\right)^2, \  \  \ B^{4i}=-q^4q^i-\left({2{\bm q\cdot k}-{\bm q}^2\over q_4}\right)(2{\bm k}^i-{\bm q}^i),\nonumber\\
&&B^{4i}+{q^4q^i\over {\bm q}^2}B^{44}=-\left({2{\bm q\cdot k}-{\bm q}^2\over q_4}\right)(2{\bm k}^i-{\bm q}^i-{2{\bm q\cdot k}-{\bm q}^2\over {\bm q}^2}q^i)=-2\left({2{\bm q\cdot k}-{\bm q}^2\over q_4}\right)\left({\bm k}^i-{{\bm q\cdot k}\over {\bm q}^2}q^i\right).
\eea
Then, apart for the ${\bm k}$-even term ${n_f^{t}({\bm k})\over E_{{\bm k}}^f}$, the left of the integrand can be evaluated alternatively by
\bea
\!\!\!\!\! {1\over 2}\left[{ B^{4i}+{q^4q^i\over {\bm q}^2}B^{44}\over 2u\ t\ E_{{\bm k}}^f( i q_4)+2{\bm q}\cdot{\bm k}+q^2}+\left.{ A^{4i}+{q^4q^i\over {\bm q}^2}A^{44}\over 2u\ t\ E_{{\bm k}}^f( i q_4)+2{\bm q}\cdot{\bm k}+q^2}\right
|_{{\bm k}\rightarrow-{\bm k}}\right]=-4{({k}^i-{{\bm q\cdot k}\over {\bm q}^2}q^i){\bm q}\cdot{\bm k}\left(2u\ t\ E_{{\bm k}}^f( i q_4)-q_4^2\right)\over (2u\ t\ E_{{\bm k}}^f( i q_4)+q^2)^2-(2{\bm q}\cdot{\bm k})^2}\label{AL}
\eea 
for the integration over ${\bm k}$. As we can see, ${\bm k}-{{\bm q\cdot k}\over {\bm q}^2}{\bm q}$ is the component of ${\bm k}$ perpendicular to ${\bm q}$ while ${\bm q}\cdot{\bm k}$ is determined by ${\bm k}$ parallel to ${\bm q}$, we can change the sign of ${\bm k}-{{\bm q\cdot k}\over {\bm q}^2}{\bm q}$ but keep that of ${\bm q}\cdot{\bm k}$ for the integration and the convolution of \eqref{AL} with ${n_f^{t}({\bm k})\over E_{{\bm k}}^f}$ must vanish. So, the relation $\Pi_{\rm \rho^0q(t)}^{4i}(q)=-{q^4q^i\over {\bm q}^2}\Pi_{\rm \rho^0q(t)}^{\rm 44}(q)$ is indeed satisfied and the Ward Identity $q_4\Pi_{ \rm \rho^0q(t)}^{\rm 44}(q)+q_i\Pi_{ \rm \rho^0q(t)}^{\rm i4}(q)=0$ is intact as should be.

Thirdly, turn to the transverse part, it can be reduced as the following,
\bea
\!\!\!\!\!&&\Pi_{ \rm \rho^0q(t)}^{ij}(q)=\!-{3g_{v q}^2\over2}g^{ij}\sum_{f=u,d}\int{d^3{\bm k}\over(2\pi)^3}\sum_{t=\pm}\!{n_f^{t}({\bm k})\over E_{{\bm k}}^f}-{3g_{v q}^2\over4}\sum_{f=u,d}\int{d^3{\bm k}\over(2\pi)^3}\sum_{u,t=\pm}{ q^2g^{ij}\!-\!q^i q^j\!+\!\left(2{\bm k}-{\bm q}\right)^i\left(2{\bm k}-{\bm q}\right)^j\over [2u\ t\ E_{{\bm k}}^f( i q_4)+2{\bm q}\cdot{\bm k}+q^2]}{n_f^{t}({\bm k})\over E_{{\bm k}}^f}\!\!\!\!\nonumber\\
\!\!\!\!\!&=&\!\!-{3g_{v q}^2\over2}g^{ij}\!\!\sum_{f=u,d}\!\int\!\!{d^3{\bm k}\over(2\pi)^3}\!\sum_{t=\pm}\!\!{n_f^{t}({\bm k})\over E_{{\bm k}}^f}-{3g_{v q}^2\over4}\!\!\sum_{f=u,d}\!\int\!\!{d^3{\bm k}\over(2\pi)^3}\!\!\sum_{u,t=\pm}\!\!\!{ q^2g^{ij}\!-\!q^i q^j\!+\!4\left({\bm k}\!-\!{{\bm q\cdot k}\over {\bm q}^2}{\bm q}\right)^i\!\left({\bm k}\!-\!{{\bm q\cdot k}\over {\bm q}^2}{\bm q}\right)^j\!\!\!+\!\left(1\!-\!2{{\bm q\cdot k}\over {\bm q}^2}\right)^2\!\!{\bm q}^i{\bm q}^j\over [2u\ t\ E_{{\bm k}}^f( i q_4)+2{\bm q}\cdot{\bm k}+q^2]}{n_f^{t}({\bm k})\over E_{{\bm k}}^f}\!\!\!\!\nonumber\\
\!\!\!\!\!&=&\!\!-{3g_{v q}^2\over2}g^{ij}\!\!\sum_{f=u,d}\!\int\!\!{d^3{\bm k}\over(2\pi)^3}\!\sum_{t=\pm}\!\!{n_f^{t}({\bm k})\over E_{{\bm k}}^f}-{3g_{v q}^2\over4}\!\!\sum_{f=u,d}\!\int\!\!{d^3{\bm k}\over(2\pi)^3}\!\!\sum_{u,t=\pm}\!\!\!{ q^2g^{ij}\!-\!q^i q^j\!-\!4\left({\bm k}\!-\!{{\bm q\cdot k}\over {\bm q}^2}{\bm q}\right)^2\!\left(g^{ij}\!+\!{q^i q^j\over  {\bm q}^2}\right)\!+\!\left(1\!-\!2{{\bm q\cdot k}\over {\bm q}^2}\right)^2\!\!{\bm q}^i{\bm q}^j\over [2u\ t\ E_{{\bm k}}^f( i q_4)+2{\bm q}\cdot{\bm k}+q^2]}{n_f^{t}({\bm k})\over E_{{\bm k}}^f}\!\!\!\!\nonumber\\
\!\!\!\!\!&=&\!\!-{3g_{v q}^2\over2}g^{ij}\!\sum_{f=u,d}\!\int\!\!{d^3{\bm k}\over(2\pi)^3}\!\sum_{t=\pm}\!{n_f^{t}({\bm k})\over E_{{\bm k}}^f}-{3g_{v q}^2\over4}\!\!\sum_{f=u,d}\!\int\!\!{d^3{\bm k}\over(2\pi)^3}\!\!\sum_{u,t=\pm}\!\!\!{ \left[q^2\!-\!4\left({\bm k}^2\!-\!{({\bm q\cdot k})^2\over {\bm q}^2}\right)\right]g^{ij}-4\left[{\bm k}^2+{\bm q\cdot k}-2{({\bm q\cdot k})^2\over {\bm q}^2}\right]{q^i q^j\over {\bm q}^2}\over [2u\ t\ E_{{\bm k}}^f( i q_4)+2{\bm q}\cdot{\bm k}+q^2]}{n_f^{t}({\bm k})\over E_{{\bm k}}^f}\!\!\!\!\nonumber\\
\!\!\!\!\!&=&\!\!{3g_{v q}^2\over2}{q_4^2\over {\bm q}^2}\left(g^{ij}+2{q^i q^j\over {\bm q}^2}\right)\sum_{f=u,d}\int{d^3{\bm k}\over(2\pi)^3}\sum_{t=\pm}{n_f^{t}({\bm k})\over E_{{\bm k}}^f}+{3g_{v q}^2\over4}\sum_{f=u,d}\int{d^3{\bm k}\over(2\pi)^3}\sum_{u,t=\pm}{1\over [2u\ t\ E_{{\bm k}}^f( i q_4)+2{\bm q}\cdot{\bm k}+q^2]}{n_f^{t}({\bm k})\over E_{{\bm k}}^f}\!\!\!\!\nonumber\\
\!\!\!\!\!&&\times\left\{ \left[q^2-4{\bm k}^2+{(2u\ t\ E_{{\bm k}}^f( i q_4)+q^2)^2\over {\bm q}^2}\right]g^{ij}+2\left[q^2-2{\bm k}^2+{(2u\ t\ E_{{\bm k}}^f( i q_4)+q^2)^2\over {\bm q}^2}+2u\ t\ E_{{\bm k}}^f( i q_4)\right]{q^i q^j\over {\bm q}^2}\right\}\!\!\!\!\nonumber\\
\!\!\!\!\!&=&\!\!{3g_{v q}^2\over2}{q_4^2\over {\bm q}^2}\left(g^{ij}+2{q^i q^j\over {\bm q}^2}\right)\sum_{f=u,d}\int{d^3{\bm k}\over(2\pi)^3}\sum_{t=\pm}{n_f^{t}({\bm k})\over E_{{\bm k}}^f}+{3g_{v q}^2\over4}\sum_{f=u,d}\int{d^3{\bm k}\over(2\pi)^3}\sum_{u,t=\pm}{1\over [2u\ t\ E_{{\bm k}}^f( i q_4)+2{\bm q}\cdot{\bm k}+q^2]}{n_f^{t}({\bm k})\over E_{{\bm k}}^f}\!\!\!\!\nonumber\\
\!\!\!\!\!&&\times\left\{\left[-{q_4^2\over {\bm q}^2}\left(4(E_{{\bm k}}^f)^2\!+\!q^2\right)-4{\bm k}^2+{q^2\over {\bm q}^2}4u\ t\ E_{{\bm k}}^f( i q_4)\right]g^{ij}\!+\!2\left[-{q_4^2\over {\bm q}^2}\left(4(E_{{\bm k}}^f)^2\!+\!q^2\right)-2{\bm k}^2+{q^2-q_4^2\over {\bm q}^2}2u\ t\ E_{{\bm k}}^f( i q_4)\right]{q^i q^j\over {\bm q}^2}\right\}\!\!\!\!\nonumber\\
\!\!\!\!\!&=&\!\!{3g_{v q}^2\over8}\!\sum_{f=u,d}\!\int\!{k\di k\over(2\pi)^2|{\bm q}|}\sum_{t=\pm}\!{n_f^{t}({\bm k})\over E_{{\bm k}}^f}\! \left\{{q_4^2\over {\bm q}^2}8{k |{\bm q}|}\left(g^{ij}\!+\!2{q^i q^j\over {\bm q}^2}\right)\!+\!\left[\left(-{q_4^2\over {\bm q}^2}\left(4(E_{{\bm k}}^f)^2\!+\!q^2\right)\!-\!4{\bm k}^2\right)\ln \alpha_f\!+\!{4q^2\over {\bm q}^2} E_{{\bm k}}^f( i q_4)\ln \beta_f\right]g^{ij}\right.\!\!\!\!\nonumber\\
\!\!\!\!\!&&\left.+2\left[\left(-{q_4^2\over {\bm q}^2}\left(4(E_{{\bm k}}^f)^2+q^2\right)-2{\bm k}^2\right)\ln \alpha_f+2{q^2-q_4^2\over {\bm q}^2}E_{{\bm k}}^f( i q_4)\ln \beta_f\right]{q^i q^j\over {\bm q}^2}\right\}.\!\!\!\!
\eea
In the third step, we have used the property $\left({\bm k}-{{\bm q\cdot k}\over {\bm q}^2}{\bm q}\right)\bot {\bm q}$ to separate the terms parallel to and perpendicular to ${\bm q}$. The tensor structure $\left(g^{ij}+{q^i q^j\over  {\bm q}^2}\right)$ is obtained by transforming the coordinates from the momentum-dependent Descartes system with axes along $ {\bm q},  {\bm q}\times  {\bm k}, \left({\bm k}-{{\bm q\cdot k}\over {\bm q}^2}{\bm q}\right)$ to the momentum-independent system with axes along $x, y, z$. Notice that the tensor structure well keeps the symmetry between the indices $i$ and $j$ and the following features
\bea
{\bm q}_i\left({\bm k}-{{\bm q\cdot k}\over {\bm q}^2}{\bm q}\right)^i=0,\ \left({\bm k}-{{\bm q\cdot k}\over {\bm q}^2}{\bm q}\right)_i\left({\bm k}-{{\bm q\cdot k}\over {\bm q}^2}{\bm q}\right)^i\left({\bm k}-{{\bm q\cdot k}\over {\bm q}^2}{\bm q}\right)^j=-\left({\bm k}-{{\bm q\cdot k}\over {\bm q}^2}{\bm q}\right)^2\left({\bm k}-{{\bm q\cdot k}\over {\bm q}^2}{\bm q}\right)^j.
\eea
Similar to the derivations in \eqref{Pi44}, we have gotten rid of the term ${\bm q}\cdot{\bm k}$ in the numerator in the fifth step and completed both the integrations over solid angles and summation over $u$ in the last step. Again, one can check that the Ward Identity $q_4\Pi_{ \rm \rho^0q(t)}^{\rm 4j}(q)+q_i\Pi_{ \rm \rho^0q(t)}^{ij}(q)=0$ is intact as should be for the spatial part with $j=1,2,3$.

Eventually, the contributions of the thermal part to the self-energies of longitudinal and transverse polarized $\rho^0$ mesons can be derived and given in Minkowski space as
\bea
\!\!\!\!\!\!F_{\rm q}^t\!\!&=&\!\!{3g_{v q}^2\over8}{M^2\over {\bm q}^2} \sum_{f=u,d}^{t=\pm}\int{k\di k\over(2\pi)^2|{\bm q}|}{n_f^{t}({\bm k})\over E_{{\bm k}}^f}\left[-8|{\bm q}|k\!+\!\left(4\left(E_{{\bm k}}^f\right)^2\!+\!M^2\right)\left(\ln |\alpha_f|\!-\!i\pi\Delta_f\right)\!+\!4 E_{{\bm k}}^fE\left(\ln |\beta_f|\!+\!i\pi\Delta_f\right)\right], \\
\!\!\!\!\!\!G_{\rm q}^t\!\!&=&\!\!{3g_{v q}^2\over8}{E^2\over {\bm q}^2}\!\sum_{f=u,d}^{t=\pm}\!\int\!\!{k\di k\over(2\pi)^2|{\bm q}|}{n_f^{t}({\bm k})\over E_{{\bm k}}^f}\!\left[8{k |{\bm q}|}\!-\!\left(4(E_{{\bm k}}^f)^2\!+\!M^2\!-\!4{|{\bm q}|^2k^2\over E^2}\right)\left(\ln |\alpha_f|\!-\!i\pi\Delta_f\right)\!-\!4{E_{{\bm k}}^f\over E}M^2\left(\ln |\beta_f|\!+\!i\pi\Delta_f\right)\right]\eea
with the auxiliary functions
\bea
&&\alpha_f\equiv{(M^2+2k|{\bm q}|)^2-4(E_{{\bm k}}^f)^2E^2\over (M^2-2k|{\bm q}|)^2-4(E_{{\bm k}}^f)^2E^2},\
\beta_f\equiv{M^4-4(k|{\bm q}|+E_{{\bm k}}^f E)^2\over M^4-4(k|{\bm q}|-E_{{\bm k}}^f E)^2},\\
&&\Delta_f={\rm Boole}(\left|E\sqrt{1\!-\!4\tilde{m}^2_f}- |{\bm q}|\right|\leq 2k\leq E\sqrt{1\!-\!4\tilde{m}^2_f}+|{\bm q}|).
\eea
The main structures of $F_{\rm q}^t$ and $G_{\rm q}^t$ are similar to those of $F_{\pi}^t$ in \eqref{Fm} and $G_{\pi}^t$ in \eqref{Gm} and can be rendered the same by taking the substitutions, $E^2\rightarrow M^2$ (thus $M^2\rightarrow E^2-2|{\bm q}|^2$) in the nonlogarithmic terms and $M^4\rightarrow E^2M^2$ in front of $\ln |\alpha_{\pi}|$.  
\end{widetext}

\bibliography{ref-lib}

\begin{thebibliography}{31}%
\makeatletter
\providecommand \@ifxundefined [1]{%
 \@ifx{#1\undefined}
}%
\providecommand \@ifnum [1]{%
 \ifnum #1\expandafter \@firstoftwo
 \else \expandafter \@secondoftwo
 \fi
}%
\providecommand \@ifx [1]{%
 \ifx #1\expandafter \@firstoftwo
 \else \expandafter \@secondoftwo
 \fi
}%
\providecommand \natexlab [1]{#1}%
\providecommand \enquote  [1]{``#1''}%
\providecommand \bibnamefont  [1]{#1}%
\providecommand \bibfnamefont [1]{#1}%
\providecommand \citenamefont [1]{#1}%
\providecommand \href@noop [0]{\@secondoftwo}%
\providecommand \href [0]{\begingroup \@sanitize@url \@href}%
\providecommand \@href[1]{\@@startlink{#1}\@@href}%
\providecommand \@@href[1]{\endgroup#1\@@endlink}%
\providecommand \@sanitize@url [0]{\catcode `\\12\catcode `\$12\catcode
  `\&12\catcode `\#12\catcode `\^12\catcode `\_12\catcode `\%12\relax}%
\providecommand \@@startlink[1]{}%
\providecommand \@@endlink[0]{}%
\providecommand \url  [0]{\begingroup\@sanitize@url \@url }%
\providecommand \@url [1]{\endgroup\@href {#1}{\urlprefix }}%
\providecommand \urlprefix  [0]{URL }%
\providecommand \Eprint [0]{\href }%
\providecommand \doibase [0]{https://doi.org/}%
\providecommand \selectlanguage [0]{\@gobble}%
\providecommand \bibinfo  [0]{\@secondoftwo}%
\providecommand \bibfield  [0]{\@secondoftwo}%
\providecommand \translation [1]{[#1]}%
\providecommand \BibitemOpen [0]{}%
\providecommand \bibitemStop [0]{}%
\providecommand \bibitemNoStop [0]{.\EOS\space}%
\providecommand \EOS [0]{\spacefactor3000\relax}%
\providecommand \BibitemShut  [1]{\csname bibitem#1\endcsname}%
\let\auto@bib@innerbib\@empty
\bibitem [{\citenamefont {Abdallah}\ \emph {et~al.}(2022)\citenamefont
  {Abdallah} \emph {et~al.}}]{STAR:2021fge}%
  \BibitemOpen
  \bibfield  {author} {\bibinfo {author} {\bibfnamefont {M.~S.}\ \bibnamefont
  {Abdallah}} \emph {et~al.} (\bibinfo {collaboration} {STAR}),\ }\bibfield
  {title} {\bibinfo {title} {{Measurements of Proton High Order Cumulants in
  $\sqrt{s_{_{\mathrm{NN}}}}$ = 3 GeV Au+Au Collisions and Implications for the
  QCD Critical Point}},\ }\href
  {https://doi.org/10.1103/PhysRevLett.128.202303} {\bibfield  {journal}
  {\bibinfo  {journal} {Phys. Rev. Lett.}\ }\textbf {\bibinfo {volume} {128}},\
  \bibinfo {pages} {202303} (\bibinfo {year} {2022})},\ \Eprint
  {https://arxiv.org/abs/2112.00240} {arXiv:2112.00240 [nucl-ex]} \BibitemShut
  {NoStop}%
\bibitem [{\citenamefont {Abdallah}\ \emph {et~al.}(2023)\citenamefont
  {Abdallah} \emph {et~al.}}]{STAR:2022etb}%
  \BibitemOpen
  \bibfield  {author} {\bibinfo {author} {\bibfnamefont {M.}~\bibnamefont
  {Abdallah}} \emph {et~al.} (\bibinfo {collaboration} {STAR}),\ }\bibfield
  {title} {\bibinfo {title} {{Higher-order cumulants and correlation functions
  of proton multiplicity distributions in sNN=3~GeV~Au+Au collisions at the
  RHIC STAR experiment}},\ }\href {https://doi.org/10.1103/PhysRevC.107.024908}
  {\bibfield  {journal} {\bibinfo  {journal} {Phys. Rev. C}\ }\textbf {\bibinfo
  {volume} {107}},\ \bibinfo {pages} {024908} (\bibinfo {year} {2023})},\
  \Eprint {https://arxiv.org/abs/2209.11940} {arXiv:2209.11940 [nucl-ex]}
  \BibitemShut {NoStop}%
\bibitem [{\citenamefont {Aboona}\ \emph {et~al.}(2025)\citenamefont {Aboona}
  \emph {et~al.}}]{STAR:2025zdq}%
  \BibitemOpen
  \bibfield  {author} {\bibinfo {author} {\bibfnamefont {B.~E.}\ \bibnamefont
  {Aboona}} \emph {et~al.} (\bibinfo {collaboration} {STAR}),\ }\bibfield
  {title} {\bibinfo {title} {{Precision Measurement of Net-Proton-Number
  Fluctuations in Au+Au Collisions at RHIC}},\ }\href
  {https://doi.org/10.1103/9l69-2d7p} {\bibfield  {journal} {\bibinfo
  {journal} {Phys. Rev. Lett.}\ }\textbf {\bibinfo {volume} {135}},\ \bibinfo
  {pages} {142301} (\bibinfo {year} {2025})},\ \Eprint
  {https://arxiv.org/abs/2504.00817} {arXiv:2504.00817 [nucl-ex]} \BibitemShut
  {NoStop}%
\bibitem [{\citenamefont {Stephanov}(2011)}]{Stephanov:2011pb}%
  \BibitemOpen
  \bibfield  {author} {\bibinfo {author} {\bibfnamefont {M.~A.}\ \bibnamefont
  {Stephanov}},\ }\bibfield  {title} {\bibinfo {title} {{On the sign of
  kurtosis near the QCD critical point}},\ }\href
  {https://doi.org/10.1103/PhysRevLett.107.052301} {\bibfield  {journal}
  {\bibinfo  {journal} {Phys. Rev. Lett.}\ }\textbf {\bibinfo {volume} {107}},\
  \bibinfo {pages} {052301} (\bibinfo {year} {2011})},\ \Eprint
  {https://arxiv.org/abs/1104.1627} {arXiv:1104.1627 [hep-ph]} \BibitemShut
  {NoStop}%
\bibitem [{\citenamefont {Fu}\ \emph {et~al.}(2021)\citenamefont {Fu},
  \citenamefont {Luo}, \citenamefont {Pawlowski}, \citenamefont {Rennecke},
  \citenamefont {Wen},\ and\ \citenamefont {Yin}}]{Fu:2021oaw}%
  \BibitemOpen
  \bibfield  {author} {\bibinfo {author} {\bibfnamefont {W.-j.}\ \bibnamefont
  {Fu}}, \bibinfo {author} {\bibfnamefont {X.}~\bibnamefont {Luo}}, \bibinfo
  {author} {\bibfnamefont {J.~M.}\ \bibnamefont {Pawlowski}}, \bibinfo {author}
  {\bibfnamefont {F.}~\bibnamefont {Rennecke}}, \bibinfo {author}
  {\bibfnamefont {R.}~\bibnamefont {Wen}},\ and\ \bibinfo {author}
  {\bibfnamefont {S.}~\bibnamefont {Yin}},\ }\bibfield  {title} {\bibinfo
  {title} {{Hyper-order baryon number fluctuations at finite temperature and
  density}},\ }\href {https://doi.org/10.1103/PhysRevD.104.094047} {\bibfield
  {journal} {\bibinfo  {journal} {Phys. Rev. D}\ }\textbf {\bibinfo {volume}
  {104}},\ \bibinfo {pages} {094047} (\bibinfo {year} {2021})},\ \Eprint
  {https://arxiv.org/abs/2101.06035} {arXiv:2101.06035 [hep-ph]} \BibitemShut
  {NoStop}%
\bibitem [{\citenamefont {Fu}\ \emph {et~al.}(2025)\citenamefont {Fu},
  \citenamefont {Luo}, \citenamefont {Pawlowski}, \citenamefont {Rennecke},\
  and\ \citenamefont {Yin}}]{Fu:2023lcm}%
  \BibitemOpen
  \bibfield  {author} {\bibinfo {author} {\bibfnamefont {W.-j.}\ \bibnamefont
  {Fu}}, \bibinfo {author} {\bibfnamefont {X.}~\bibnamefont {Luo}}, \bibinfo
  {author} {\bibfnamefont {J.~M.}\ \bibnamefont {Pawlowski}}, \bibinfo {author}
  {\bibfnamefont {F.}~\bibnamefont {Rennecke}},\ and\ \bibinfo {author}
  {\bibfnamefont {S.}~\bibnamefont {Yin}},\ }\bibfield  {title} {\bibinfo
  {title} {{Ripples of the QCD critical point}},\ }\href
  {https://doi.org/10.1103/PhysRevD.111.L031502} {\bibfield  {journal}
  {\bibinfo  {journal} {Phys. Rev. D}\ }\textbf {\bibinfo {volume} {111}},\
  \bibinfo {pages} {L031502} (\bibinfo {year} {2025})},\ \Eprint
  {https://arxiv.org/abs/2308.15508} {arXiv:2308.15508 [hep-ph]} \BibitemShut
  {NoStop}%
\bibitem [{\citenamefont {Lu}\ \emph {et~al.}(2025)\citenamefont {Lu},
  \citenamefont {Gao}, \citenamefont {Liu},\ and\ \citenamefont
  {Pawlowski}}]{Lu:2025cls}%
  \BibitemOpen
  \bibfield  {author} {\bibinfo {author} {\bibfnamefont {Y.}~\bibnamefont
  {Lu}}, \bibinfo {author} {\bibfnamefont {F.}~\bibnamefont {Gao}}, \bibinfo
  {author} {\bibfnamefont {Y.-x.}\ \bibnamefont {Liu}},\ and\ \bibinfo {author}
  {\bibfnamefont {J.~M.}\ \bibnamefont {Pawlowski}},\ }\bibfield  {title}
  {\bibinfo {title} {{Finite density signatures of confining and chiral
  dynamics in QCD thermodynamics and fluctuations of conserved charges}},\
  }\href@noop {} {\  (\bibinfo {year} {2025})},\ \Eprint
  {https://arxiv.org/abs/2504.05099} {arXiv:2504.05099 [hep-ph]} \BibitemShut
  {NoStop}%
\bibitem [{\citenamefont {Luo}\ and\ \citenamefont {Xu}(2017)}]{Luo:2017faz}%
  \BibitemOpen
  \bibfield  {author} {\bibinfo {author} {\bibfnamefont {X.}~\bibnamefont
  {Luo}}\ and\ \bibinfo {author} {\bibfnamefont {N.}~\bibnamefont {Xu}},\
  }\bibfield  {title} {\bibinfo {title} {{Search for the QCD Critical Point
  with Fluctuations of Conserved Quantities in Relativistic Heavy-Ion
  Collisions at RHIC : An Overview}},\ }\href
  {https://doi.org/10.1007/s41365-017-0257-0} {\bibfield  {journal} {\bibinfo
  {journal} {Nucl. Sci. Tech.}\ }\textbf {\bibinfo {volume} {28}},\ \bibinfo
  {pages} {112} (\bibinfo {year} {2017})},\ \Eprint
  {https://arxiv.org/abs/1701.02105} {arXiv:1701.02105 [nucl-ex]} \BibitemShut
  {NoStop}%
\bibitem [{\citenamefont {Luo}\ \emph {et~al.}(2022)\citenamefont {Luo},
  \citenamefont {Wang}, \citenamefont {Xu},\ and\ \citenamefont
  {Zhuang}}]{Luo:2022mtp}%
  \BibitemOpen
  \bibinfo {editor} {\bibfnamefont {X.}~\bibnamefont {Luo}}, \bibinfo {editor}
  {\bibfnamefont {Q.}~\bibnamefont {Wang}}, \bibinfo {editor} {\bibfnamefont
  {N.}~\bibnamefont {Xu}},\ and\ \bibinfo {editor} {\bibfnamefont
  {P.}~\bibnamefont {Zhuang}},\ eds.,\ \href
  {https://doi.org/10.1007/978-981-19-4441-3} {\emph {\bibinfo {title}
  {{Properties of QCD Matter at High Baryon Density}}}}\ (\bibinfo  {publisher}
  {Springer},\ \bibinfo {year} {2022})\BibitemShut {NoStop}%
\bibitem [{\citenamefont {Dupuis}\ \emph {et~al.}(2021)\citenamefont {Dupuis},
  \citenamefont {Canet}, \citenamefont {Eichhorn}, \citenamefont {Metzner},
  \citenamefont {Pawlowski}, \citenamefont {Tissier},\ and\ \citenamefont
  {Wschebor}}]{Dupuis:2020fhh}%
  \BibitemOpen
  \bibfield  {author} {\bibinfo {author} {\bibfnamefont {N.}~\bibnamefont
  {Dupuis}}, \bibinfo {author} {\bibfnamefont {L.}~\bibnamefont {Canet}},
  \bibinfo {author} {\bibfnamefont {A.}~\bibnamefont {Eichhorn}}, \bibinfo
  {author} {\bibfnamefont {W.}~\bibnamefont {Metzner}}, \bibinfo {author}
  {\bibfnamefont {J.~M.}\ \bibnamefont {Pawlowski}}, \bibinfo {author}
  {\bibfnamefont {M.}~\bibnamefont {Tissier}},\ and\ \bibinfo {author}
  {\bibfnamefont {N.}~\bibnamefont {Wschebor}},\ }\bibfield  {title} {\bibinfo
  {title} {{The nonperturbative functional renormalization group and its
  applications}},\ }\href {https://doi.org/10.1016/j.physrep.2021.01.001}
  {\bibfield  {journal} {\bibinfo  {journal} {Phys. Rept.}\ }\textbf {\bibinfo
  {volume} {910}},\ \bibinfo {pages} {1} (\bibinfo {year} {2021})},\ \Eprint
  {https://arxiv.org/abs/2006.04853} {arXiv:2006.04853 [cond-mat.stat-mech]}
  \BibitemShut {NoStop}%
\bibitem [{\citenamefont {Fu}(2022)}]{Fu:2022gou}%
  \BibitemOpen
  \bibfield  {author} {\bibinfo {author} {\bibfnamefont {W.-j.}\ \bibnamefont
  {Fu}},\ }\bibfield  {title} {\bibinfo {title} {{QCD at finite temperature and
  density within the fRG approach: an overview}},\ }\href
  {https://doi.org/10.1088/1572-9494/ac86be} {\bibfield  {journal} {\bibinfo
  {journal} {Commun. Theor. Phys.}\ }\textbf {\bibinfo {volume} {74}},\
  \bibinfo {pages} {097304} (\bibinfo {year} {2022})},\ \Eprint
  {https://arxiv.org/abs/2205.00468} {arXiv:2205.00468 [hep-ph]} \BibitemShut
  {NoStop}%
\bibitem [{\citenamefont {McLerran}\ and\ \citenamefont
  {Toimela}(1985)}]{McLerran:1984ay}%
  \BibitemOpen
  \bibfield  {author} {\bibinfo {author} {\bibfnamefont {L.~D.}\ \bibnamefont
  {McLerran}}\ and\ \bibinfo {author} {\bibfnamefont {T.}~\bibnamefont
  {Toimela}},\ }\bibfield  {title} {\bibinfo {title} {{Photon and Dilepton
  Emission from the Quark - Gluon Plasma: Some General Considerations}},\
  }\href {https://doi.org/10.1103/PhysRevD.31.545} {\bibfield  {journal}
  {\bibinfo  {journal} {Phys. Rev. D}\ }\textbf {\bibinfo {volume} {31}},\
  \bibinfo {pages} {545} (\bibinfo {year} {1985})}\BibitemShut {NoStop}%
\bibitem [{\citenamefont {Rapp}\ and\ \citenamefont
  {Wambach}(2000)}]{Rapp:1999ej}%
  \BibitemOpen
  \bibfield  {author} {\bibinfo {author} {\bibfnamefont {R.}~\bibnamefont
  {Rapp}}\ and\ \bibinfo {author} {\bibfnamefont {J.}~\bibnamefont {Wambach}},\
  }\bibfield  {title} {\bibinfo {title} {{Chiral symmetry restoration and
  dileptons in relativistic heavy ion collisions}},\ }\href
  {https://doi.org/10.1007/0-306-47101-9_1} {\bibfield  {journal} {\bibinfo
  {journal} {Adv. Nucl. Phys.}\ }\textbf {\bibinfo {volume} {25}},\ \bibinfo
  {pages} {1} (\bibinfo {year} {2000})},\ \Eprint
  {https://arxiv.org/abs/hep-ph/9909229} {arXiv:hep-ph/9909229} \BibitemShut
  {NoStop}%
\bibitem [{\citenamefont {Stephanov}(2009)}]{Stephanov:2008qz}%
  \BibitemOpen
  \bibfield  {author} {\bibinfo {author} {\bibfnamefont {M.}~\bibnamefont
  {Stephanov}},\ }\bibfield  {title} {\bibinfo {title} {{Non-Gaussian
  fluctuations near the QCD critical point}},\ }\href
  {https://doi.org/10.1103/PhysRevLett.102.032301} {\bibfield  {journal}
  {\bibinfo  {journal} {Phys. Rev. Lett.}\ }\textbf {\bibinfo {volume} {102}},\
  \bibinfo {pages} {032301} (\bibinfo {year} {2009})},\ \Eprint
  {https://arxiv.org/abs/0809.3450} {arXiv:0809.3450 [hep-ph]} \BibitemShut
  {NoStop}%
\bibitem [{\citenamefont {McLerran}(1987)}]{McLerran:1987pz}%
  \BibitemOpen
  \bibfield  {author} {\bibinfo {author} {\bibfnamefont {L.~D.}\ \bibnamefont
  {McLerran}},\ }\bibfield  {title} {\bibinfo {title} {{A Chiral Symmetry Order
  Parameter, the Lattice and Nucleosynthesis}},\ }\href
  {https://doi.org/10.1103/PhysRevD.36.3291} {\bibfield  {journal} {\bibinfo
  {journal} {Phys. Rev. D}\ }\textbf {\bibinfo {volume} {36}},\ \bibinfo
  {pages} {3291} (\bibinfo {year} {1987})}\BibitemShut {NoStop}%
\bibitem [{\citenamefont {Schaefer}\ \emph {et~al.}(2010)\citenamefont
  {Schaefer}, \citenamefont {Wagner},\ and\ \citenamefont
  {Wambach}}]{Schaefer:2009ui}%
  \BibitemOpen
  \bibfield  {author} {\bibinfo {author} {\bibfnamefont {B.-J.}\ \bibnamefont
  {Schaefer}}, \bibinfo {author} {\bibfnamefont {M.}~\bibnamefont {Wagner}},\
  and\ \bibinfo {author} {\bibfnamefont {J.}~\bibnamefont {Wambach}},\
  }\bibfield  {title} {\bibinfo {title} {{Thermodynamics of (2+1)-flavor QCD:
  Confronting Models with Lattice Studies}},\ }\href
  {https://doi.org/10.1103/PhysRevD.81.074013} {\bibfield  {journal} {\bibinfo
  {journal} {Phys. Rev. D}\ }\textbf {\bibinfo {volume} {81}},\ \bibinfo
  {pages} {074013} (\bibinfo {year} {2010})},\ \Eprint
  {https://arxiv.org/abs/0910.5628} {arXiv:0910.5628 [hep-ph]} \BibitemShut
  {NoStop}%
\bibitem [{\citenamefont {Cao}(2025)}]{Cao:2025zvh}%
  \BibitemOpen
  \bibfield  {author} {\bibinfo {author} {\bibfnamefont {G.}~\bibnamefont
  {Cao}},\ }\bibfield  {title} {\bibinfo {title} {{Moat regimes within a
  (2+1)-flavor Polyakov-quark-meson model}},\ }\href
  {https://doi.org/10.1103/jjs8-mj6f} {\bibfield  {journal} {\bibinfo
  {journal} {Phys. Rev. D}\ }\textbf {\bibinfo {volume} {112}},\ \bibinfo
  {pages} {034013} (\bibinfo {year} {2025})},\ \Eprint
  {https://arxiv.org/abs/2504.18874} {arXiv:2504.18874 [hep-ph]} \BibitemShut
  {NoStop}%
\bibitem [{\citenamefont {Pisarski}(1995)}]{Pisarski:1995xu}%
  \BibitemOpen
  \bibfield  {author} {\bibinfo {author} {\bibfnamefont {R.~D.}\ \bibnamefont
  {Pisarski}},\ }\bibfield  {title} {\bibinfo {title} {{Where does the rho go?
  Chirally symmetric vector mesons in the quark - gluon plasma}},\ }\href
  {https://doi.org/10.1103/PhysRevD.52.R3773} {\bibfield  {journal} {\bibinfo
  {journal} {Phys. Rev. D}\ }\textbf {\bibinfo {volume} {52}},\ \bibinfo
  {pages} {R3773} (\bibinfo {year} {1995})},\ \Eprint
  {https://arxiv.org/abs/hep-ph/9503328} {arXiv:hep-ph/9503328} \BibitemShut
  {NoStop}%
\bibitem [{\citenamefont {Bellwied}\ \emph {et~al.}(2015)\citenamefont
  {Bellwied}, \citenamefont {Borsanyi}, \citenamefont {Fodor}, \citenamefont
  {Guenther}, \citenamefont {Katz}, \citenamefont {Ratti},\ and\ \citenamefont
  {Szabo}}]{Bellwied:2015rza}%
  \BibitemOpen
  \bibfield  {author} {\bibinfo {author} {\bibfnamefont {R.}~\bibnamefont
  {Bellwied}}, \bibinfo {author} {\bibfnamefont {S.}~\bibnamefont {Borsanyi}},
  \bibinfo {author} {\bibfnamefont {Z.}~\bibnamefont {Fodor}}, \bibinfo
  {author} {\bibfnamefont {J.}~\bibnamefont {Guenther}}, \bibinfo {author}
  {\bibfnamefont {S.~D.}\ \bibnamefont {Katz}}, \bibinfo {author}
  {\bibfnamefont {C.}~\bibnamefont {Ratti}},\ and\ \bibinfo {author}
  {\bibfnamefont {K.~K.}\ \bibnamefont {Szabo}},\ }\bibfield  {title} {\bibinfo
  {title} {{The QCD phase diagram from analytic continuation}},\ }\href
  {https://doi.org/10.1016/j.physletb.2015.11.011} {\bibfield  {journal}
  {\bibinfo  {journal} {Phys. Lett. B}\ }\textbf {\bibinfo {volume} {751}},\
  \bibinfo {pages} {559} (\bibinfo {year} {2015})},\ \Eprint
  {https://arxiv.org/abs/1507.07510} {arXiv:1507.07510 [hep-lat]} \BibitemShut
  {NoStop}%
\bibitem [{\citenamefont {Bazavov}\ \emph
  {et~al.}(2019{\natexlab{a}})\citenamefont {Bazavov} \emph
  {et~al.}}]{HotQCD:2018pds}%
  \BibitemOpen
  \bibfield  {author} {\bibinfo {author} {\bibfnamefont {A.}~\bibnamefont
  {Bazavov}} \emph {et~al.} (\bibinfo {collaboration} {HotQCD}),\ }\bibfield
  {title} {\bibinfo {title} {{Chiral crossover in QCD at zero and non-zero
  chemical potentials}},\ }\href
  {https://doi.org/10.1016/j.physletb.2019.05.013} {\bibfield  {journal}
  {\bibinfo  {journal} {Phys. Lett. B}\ }\textbf {\bibinfo {volume} {795}},\
  \bibinfo {pages} {15} (\bibinfo {year} {2019}{\natexlab{a}})},\ \Eprint
  {https://arxiv.org/abs/1812.08235} {arXiv:1812.08235 [hep-lat]} \BibitemShut
  {NoStop}%
\bibitem [{\citenamefont {Fu}\ \emph {et~al.}(2020)\citenamefont {Fu},
  \citenamefont {Pawlowski},\ and\ \citenamefont {Rennecke}}]{Fu:2019hdw}%
  \BibitemOpen
  \bibfield  {author} {\bibinfo {author} {\bibfnamefont {W.-j.}\ \bibnamefont
  {Fu}}, \bibinfo {author} {\bibfnamefont {J.~M.}\ \bibnamefont {Pawlowski}},\
  and\ \bibinfo {author} {\bibfnamefont {F.}~\bibnamefont {Rennecke}},\
  }\bibfield  {title} {\bibinfo {title} {{QCD phase structure at finite
  temperature and density}},\ }\href
  {https://doi.org/10.1103/PhysRevD.101.054032} {\bibfield  {journal} {\bibinfo
   {journal} {Phys. Rev. D}\ }\textbf {\bibinfo {volume} {101}},\ \bibinfo
  {pages} {054032} (\bibinfo {year} {2020})},\ \Eprint
  {https://arxiv.org/abs/1909.02991} {arXiv:1909.02991 [hep-ph]} \BibitemShut
  {NoStop}%
\bibitem [{\citenamefont {Gao}\ and\ \citenamefont
  {Pawlowski}(2021)}]{Gao:2020fbl}%
  \BibitemOpen
  \bibfield  {author} {\bibinfo {author} {\bibfnamefont {F.}~\bibnamefont
  {Gao}}\ and\ \bibinfo {author} {\bibfnamefont {J.~M.}\ \bibnamefont
  {Pawlowski}},\ }\bibfield  {title} {\bibinfo {title} {{Chiral phase structure
  and critical end point in QCD}},\ }\href
  {https://doi.org/10.1016/j.physletb.2021.136584} {\bibfield  {journal}
  {\bibinfo  {journal} {Phys. Lett. B}\ }\textbf {\bibinfo {volume} {820}},\
  \bibinfo {pages} {136584} (\bibinfo {year} {2021})},\ \Eprint
  {https://arxiv.org/abs/2010.13705} {arXiv:2010.13705 [hep-ph]} \BibitemShut
  {NoStop}%
\bibitem [{\citenamefont {Gunkel}\ and\ \citenamefont
  {Fischer}(2021)}]{Gunkel:2021oya}%
  \BibitemOpen
  \bibfield  {author} {\bibinfo {author} {\bibfnamefont {P.~J.}\ \bibnamefont
  {Gunkel}}\ and\ \bibinfo {author} {\bibfnamefont {C.~S.}\ \bibnamefont
  {Fischer}},\ }\bibfield  {title} {\bibinfo {title} {{Locating the critical
  endpoint of QCD: Mesonic backcoupling effects}},\ }\href
  {https://doi.org/10.1103/PhysRevD.104.054022} {\bibfield  {journal} {\bibinfo
   {journal} {Phys. Rev. D}\ }\textbf {\bibinfo {volume} {104}},\ \bibinfo
  {pages} {054022} (\bibinfo {year} {2021})},\ \Eprint
  {https://arxiv.org/abs/2106.08356} {arXiv:2106.08356 [hep-ph]} \BibitemShut
  {NoStop}%
\bibitem [{\citenamefont {Ratti}\ \emph {et~al.}(2006)\citenamefont {Ratti},
  \citenamefont {Thaler},\ and\ \citenamefont {Weise}}]{Ratti:2005jh}%
  \BibitemOpen
  \bibfield  {author} {\bibinfo {author} {\bibfnamefont {C.}~\bibnamefont
  {Ratti}}, \bibinfo {author} {\bibfnamefont {M.~A.}\ \bibnamefont {Thaler}},\
  and\ \bibinfo {author} {\bibfnamefont {W.}~\bibnamefont {Weise}},\ }\bibfield
   {title} {\bibinfo {title} {{Phases of QCD: Lattice thermodynamics and a
  field theoretical model}},\ }\href
  {https://doi.org/10.1103/PhysRevD.73.014019} {\bibfield  {journal} {\bibinfo
  {journal} {Phys. Rev.}\ }\textbf {\bibinfo {volume} {D73}},\ \bibinfo {pages}
  {014019} (\bibinfo {year} {2006})},\ \Eprint
  {https://arxiv.org/abs/hep-ph/0506234} {arXiv:hep-ph/0506234 [hep-ph]}
  \BibitemShut {NoStop}%
\bibitem [{\citenamefont {Herbst}\ \emph {et~al.}(2011)\citenamefont {Herbst},
  \citenamefont {Pawlowski},\ and\ \citenamefont {Schaefer}}]{Herbst:2010rf}%
  \BibitemOpen
  \bibfield  {author} {\bibinfo {author} {\bibfnamefont {T.~K.}\ \bibnamefont
  {Herbst}}, \bibinfo {author} {\bibfnamefont {J.~M.}\ \bibnamefont
  {Pawlowski}},\ and\ \bibinfo {author} {\bibfnamefont {B.-J.}\ \bibnamefont
  {Schaefer}},\ }\bibfield  {title} {\bibinfo {title} {{The phase structure of
  the Polyakovquarkmeson model beyond mean field}},\ }\href
  {https://doi.org/10.1016/j.physletb.2010.12.003} {\bibfield  {journal}
  {\bibinfo  {journal} {Phys. Lett. B}\ }\textbf {\bibinfo {volume} {696}},\
  \bibinfo {pages} {58} (\bibinfo {year} {2011})},\ \Eprint
  {https://arxiv.org/abs/1008.0081} {arXiv:1008.0081 [hep-ph]} \BibitemShut
  {NoStop}%
\bibitem [{\citenamefont {Gounaris}\ and\ \citenamefont
  {Sakurai}(1968)}]{Gounaris:1968mw}%
  \BibitemOpen
  \bibfield  {author} {\bibinfo {author} {\bibfnamefont {G.~J.}\ \bibnamefont
  {Gounaris}}\ and\ \bibinfo {author} {\bibfnamefont {J.~J.}\ \bibnamefont
  {Sakurai}},\ }\bibfield  {title} {\bibinfo {title} {{Finite width corrections
  to the vector meson dominance prediction for $\rho \to e^+ e^-$}},\ }\href
  {https://doi.org/10.1103/PhysRevLett.21.244} {\bibfield  {journal} {\bibinfo
  {journal} {Phys. Rev. Lett.}\ }\textbf {\bibinfo {volume} {21}},\ \bibinfo
  {pages} {244} (\bibinfo {year} {1968})}\BibitemShut {NoStop}%
\bibitem [{\citenamefont {Gale}\ and\ \citenamefont
  {Kapusta}(1991)}]{Gale:1990pn}%
  \BibitemOpen
  \bibfield  {author} {\bibinfo {author} {\bibfnamefont {C.}~\bibnamefont
  {Gale}}\ and\ \bibinfo {author} {\bibfnamefont {J.~I.}\ \bibnamefont
  {Kapusta}},\ }\bibfield  {title} {\bibinfo {title} {{Vector dominance model
  at finite temperature}},\ }\href
  {https://doi.org/10.1016/0550-3213(91)90459-B} {\bibfield  {journal}
  {\bibinfo  {journal} {Nucl. Phys. B}\ }\textbf {\bibinfo {volume} {357}},\
  \bibinfo {pages} {65} (\bibinfo {year} {1991})}\BibitemShut {NoStop}%
\bibitem [{\citenamefont {Kajantie}\ \emph {et~al.}(1986)\citenamefont
  {Kajantie}, \citenamefont {Kapusta}, \citenamefont {McLerran},\ and\
  \citenamefont {Mekjian}}]{Kajantie:1986dh}%
  \BibitemOpen
  \bibfield  {author} {\bibinfo {author} {\bibfnamefont {K.}~\bibnamefont
  {Kajantie}}, \bibinfo {author} {\bibfnamefont {J.~I.}\ \bibnamefont
  {Kapusta}}, \bibinfo {author} {\bibfnamefont {L.~D.}\ \bibnamefont
  {McLerran}},\ and\ \bibinfo {author} {\bibfnamefont {A.}~\bibnamefont
  {Mekjian}},\ }\bibfield  {title} {\bibinfo {title} {{Dilepton Emission and
  the QCD Phase Transition in Ultrarelativistic Nuclear Collisions}},\ }\href
  {https://doi.org/10.1103/PhysRevD.34.2746} {\bibfield  {journal} {\bibinfo
  {journal} {Phys. Rev. D}\ }\textbf {\bibinfo {volume} {34}},\ \bibinfo
  {pages} {2746} (\bibinfo {year} {1986})}\BibitemShut {NoStop}%
\bibitem [{\citenamefont {Bazavov}\ \emph
  {et~al.}(2019{\natexlab{b}})\citenamefont {Bazavov} \emph
  {et~al.}}]{Bazavov:2019www}%
  \BibitemOpen
  \bibfield  {author} {\bibinfo {author} {\bibfnamefont {A.}~\bibnamefont
  {Bazavov}} \emph {et~al.},\ }\bibfield  {title} {\bibinfo {title} {{Meson
  screening masses in (2+1)-flavor QCD}},\ }\href
  {https://doi.org/10.1103/PhysRevD.100.094510} {\bibfield  {journal} {\bibinfo
   {journal} {Phys. Rev. D}\ }\textbf {\bibinfo {volume} {100}},\ \bibinfo
  {pages} {094510} (\bibinfo {year} {2019}{\natexlab{b}})},\ \Eprint
  {https://arxiv.org/abs/1908.09552} {arXiv:1908.09552 [hep-lat]} \BibitemShut
  {NoStop}%
\bibitem [{\citenamefont {Fukushima}(2004)}]{Fukushima:2003fw}%
  \BibitemOpen
  \bibfield  {author} {\bibinfo {author} {\bibfnamefont {K.}~\bibnamefont
  {Fukushima}},\ }\bibfield  {title} {\bibinfo {title} {{Chiral effective model
  with the Polyakov loop}},\ }\href
  {https://doi.org/10.1016/j.physletb.2004.04.027} {\bibfield  {journal}
  {\bibinfo  {journal} {Phys. Lett. B}\ }\textbf {\bibinfo {volume} {591}},\
  \bibinfo {pages} {277} (\bibinfo {year} {2004})},\ \Eprint
  {https://arxiv.org/abs/hep-ph/0310121} {arXiv:hep-ph/0310121 [hep-ph]}
  \BibitemShut {NoStop}%
\bibitem [{\citenamefont {Han}(2024)}]{Han:2024nzr}%
  \BibitemOpen
  \bibfield  {author} {\bibinfo {author} {\bibfnamefont {Y.}~\bibnamefont
  {Han}} (\bibinfo {collaboration} {STAR}),\ }\bibfield  {title} {\bibinfo
  {title} {{Thermal dielectron measurements in Au+Au collisions at
  $\sqrt{s_{NN}}$ = 7.7, 14.6, and 19.6 GeV with the STAR experiment}},\ }\href
  {https://doi.org/10.1051/epjconf/202429607004} {\bibfield  {journal}
  {\bibinfo  {journal} {EPJ Web Conf.}\ }\textbf {\bibinfo {volume} {296}},\
  \bibinfo {pages} {07004} (\bibinfo {year} {2024})}\BibitemShut {NoStop}%
\end{thebibliography}%

\end{document}